\documentclass[%
 reprint,
 amsmath,amssymb,
prb,
nofootinbib]{revtex4-2}
\usepackage{graphicx}
\usepackage{dcolumn}
\usepackage{bm}

\usepackage[english]{babel}
\usepackage{xspace}
\usepackage{float}

\newcommand{\isotope}[3]{\ensuremath{^{#1}\mathrm{#2}^{#3}}}
\newcommand{\Pb}{\isotope{208}{Pb}{82+}} 
 
\newcommand{\qty}[2]{\ensuremath{#1\,\mathrm{#2}}}  
\newcommand{\sixtrack}{SixTrack\xspace}
\newcommand{\bs}{$\beta^*$}
\def\PD#1#2{\frac{\partial #1}{\partial #2}}
\RequirePackage{subfig}
\usepackage{url}

\begin{document}


\title{Simulations of heavy-ion halo collimation at the CERN Large Hadron Collider: benchmark with measurements and cleaning performance evaluation}%
\thanks{Work partially supported by the High-Luminosity Large Hadron Collider project.}

\author{N.~Fuster-Mart\'inez } \thanks{nuria.fuster.martinez@cern.ch} 
\author{R.~Bruce}
\author{F.~Cerutti}
\author{R.~De~Maria}
\author{P.~Hermes}
\author{A. Lechner}
\author{A.~Mereghetti}
\author{J.~Molson}
\author{S.~Redaelli}
\author{E. Skordis}
\affiliation{%
CERN, Geneva, Switzerland}%

\author{A.~Abramov}
\author{L.~Nevay}
\affiliation{John Adams Institute at Royal Holloway, University of London, Egham, TW20 0EX, UK}

\date{\today}

\begin{abstract}
Protons and heavy-ion beams at unprecedented energies are brought into collisions in the CERN Large Hadron Collider for high-energy experiments. The LHC multi-stage collimation system is designed to provide protection against regular and abnormal losses in order to reduce the risk of quenches of the superconducting magnets as well as keeping background in the experiments under control. Compared to protons, beam collimation in the heavy-ion runs is more challenging despite the lower stored beam energies, because the efficiency of cleaning with heavy ions has been observed to be two orders of magnitude worse. This is due to the differences in the interaction mechanisms between the beams and the collimators. Ion beams experience fragmentation and electromagnetic dissociation at the collimators that result in a substantial flux of off-rigidity particles that escape the collimation system. These out-scattered nuclei might be lost around the ring, eventually imposing a limit on the maximum achievable stored beam energy. The more stringent limit comes from potential quenches of superconducting magnets. Accurate simulation tools are crucial in order to understand and control these losses. A new simulation framework has been developed for heavy-ion collimation based on the coupling of the \sixtrack tracking code, which has been extended to track arbitrary heavy-ion species, and the FLUKA Monte Carlo code that models the electromagnetic and nuclear interactions of the heavy-ions with the nuclei of the collimator material. In this paper, the functionality of the new simulation tool is described. Furthermore, \sixtrack-FLUKA coupling simulations are presented and compared with measurements done with \Pb~ions in the LHC. The agreement between simulations and measurements is discussed and the results are used to understand and optimise losses. The simulation tool is also applied to predict the performance of the collimation system for the High-Luminosity LHC. Based on the simulation results and the experience gained in past heavy-ion runs, some conclusions are presented.
\end{abstract}

\maketitle


\section{\label{sec:level1}Introduction}

At the CERN Large Hadron Collider (LHC)~\cite{Brüning:LHC}, proton and heavy-ion
beams are brought into collision for high-energy physics experiments. Unavoidable losses occur in colliders due to the interaction of the main beam with residual gas in the beam pipe, the collision of the beams at the interaction points, instabilities, resonances, or due to the diffusion mechanisms driven by electron cloud and beam-beam interactions, just to name a few. The LHC multi-stage collimation system~\cite{LHCcoll:1,LHCcoll:2,LHCcoll:3,LHCcoll:4,LHCcoll:5} is designed to protect the LHC hardware against regular and abnormal beam losses. In particular, the collimation system has to protect the superconducting magnets that risk quenching, changing their state from superconducting to normal conducting, reducing as a consequence the available time for physics data acquisition. In addition, the collimation system has also to keep the background in the experiments under control~\cite{bruce13_NIM_backgrounds,bruce19_PRAB_beam-halo_backgrounds_ATLAS}. 
For heavy-ion beams a degraded collimation cleaning efficiency is expected. This is due to the nuclear fragmentation and electromagnetic dissociation (EMD) processes occurring at the collimators that generate a large spectrum of secondary nuclei with a different charge-to-mass ratio with respect to the main beam. Some of these fragments can escape the downstream collimation stages and be lost at other locations around the ring~\cite{braun04,hermes_nim,braun04nr2,Redaelli:2646800}. In the LHC heavy-ion runs, a reduction by a factor 100 of the collimation cleaning efficiency has been observed in comparison to protons, which is not fully compensated by the lower stored beam energies reached during these runs~\cite{epac2004, Hermes:thesis, NFusterMartinez-EVIAN2019,jowett19_evian}. This makes the collimation of heavy-ions, which typically are operated during 1~month per year, more challenging.
\begin{figure*}[htb!]
    \centering
    \includegraphics*[width=15cm]{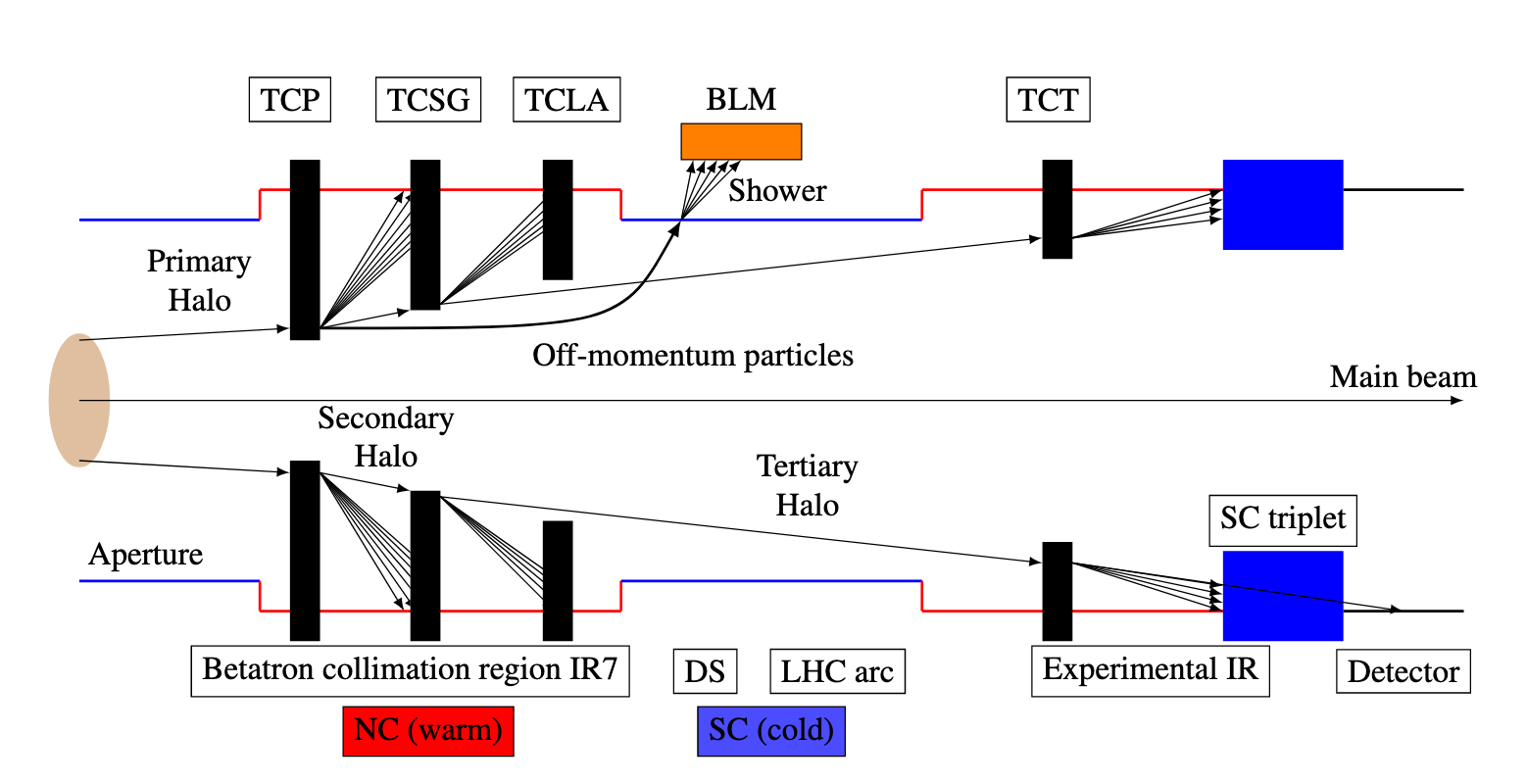}
    \caption{LHC multi-stage collimation system scheme where the different collimator families are indicated. Normal conducting (NC) and superconducting (SC) apertures are depicted in red and blue, respectively~\cite{Redaelli:2646800}.}
    \label{Coll_scheme} 
\end{figure*}

In 2018, \Pb~ion beams were accelerated to an energy of \qty{6.37\,Z}{TeV}\footnote{$Z$ is the charge number of the nuclei in the main beam.}  in the LHC~\cite{ipac19:IONRUN}. The stored beam energy reached by the ions was \qty{13.3}{MJ}, which is well above the design value of \qty{3.8}{MJ}~\cite{Brüning:LHC} and what was previously achieved at the LHC~\cite{ion2015,NFusterMartinez-EVIAN2019}. In the 2018 \Pb~ion run, no magnet quenches were recorded due to collimation losses from circulating beams, but 7 out of 48 fills were dumped due to high losses in collimators caused by orbit oscillations~\cite{Daniele_Evian}. These beam dumps could be avoided by a better collimation cleaning efficiency. This underlines the need for a solution for future runs where even higher intensities are envisaged by the High-Luminosity LHC project (HL-LHC)~\cite{Apollinari:HL-LHC}. 

The development of accurate simulation tools is crucial to understand and control the secondary fragments generated at the collimators, quantify possible limitations, and elaborate mitigation strategies for future runs. Significant progress has been made in the past years to improve the accuracy of the heavy-ion collimation simulation tools. In this paper, a new heavy-ion simulation tool developed based on the coupling of the \sixtrack tracking code ~\cite{sixtrack,sixtrack-web} and the FLUKA~\cite{fluka1,fluka14} Monte-Carlo program, similar to the development of simulation tools for protons~\cite{mereghetti13_ipac,skordis18_tracking_workshop} is presented. In addition, the performance of the LHC collimation system, simulated with the \sixtrack-FLUKA coupling tool, is presented and compared to measurements for different scenarios for a better understanding and optimization of losses in the LHC. Furthermore, using this new tool, the performance in future configurations is predicted. 

The paper is organised as follows. First, the LHC collimation system is described in Section~\ref{sec:level2} as well as the measurements performed at the start of every run to qualify the performance of the collimation system. These measurements are later used for the comparison with simulations. In Section~\ref{sec:simulation_method} the heavy-ion collimation simulation state-of-the-art tools are presented and the functionality of the \sixtrack-FLUKA coupling framework is described. In Section~\ref{sec:level3}, the methodology followed to perform the analysis of the simulations and the simulation set-up are described. In Section~\ref{sec:benchmark}, \sixtrack-FLUKA coupling simulations are presented and compared to measurements from the 2018 \Pb~ion run, and operational applications are discussed. In addition, detailed simulations performed with FLUKA and more complete geometries are compared to measurements performed in the 2015 \Pb~ion run and the agreement is discussed. In the last section, the simulation tool is used to predict the cleaning performance of the collimation system for the future upgraded HL-LHC configuration. 

\section{\label{sec:level2}Heavy-Ion Beam Collimation at the LHC}

The LHC multi-stage collimation system is organised in a well-defined hierarchy, based on their opening, with different collimator families as illustrated in Fig.~\ref{Coll_scheme}, where each individual collimator consists of two movable jaws with the beam passing in the centre. The preservation of the hierarchy between families is a pre-requisite to ensure a good performance of the system.

\begin{figure}[!htb]
    \centering
    \includegraphics*[width=8.6cm]{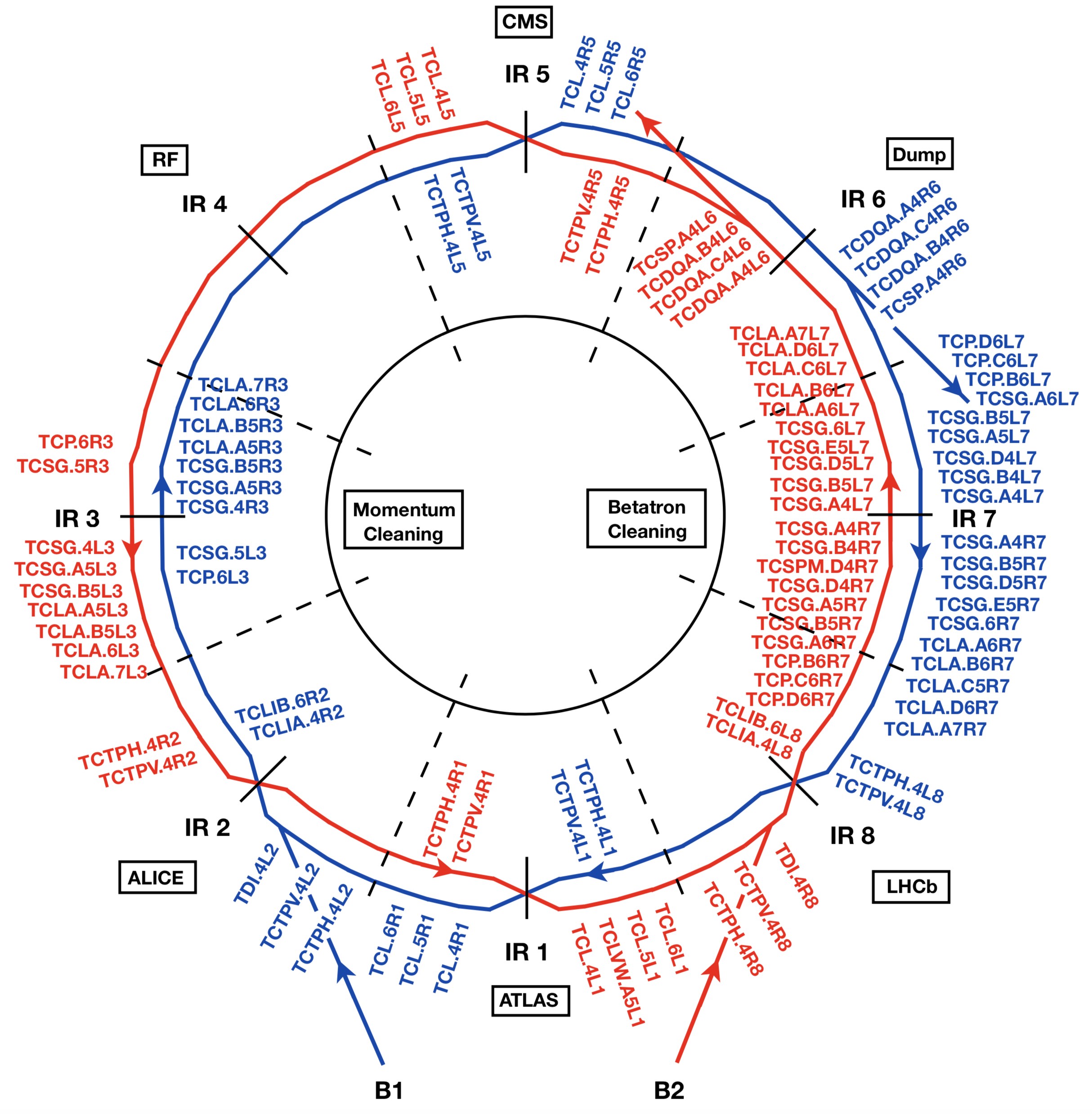}
    \caption{2018 LHC collimation system layout~\cite{Gabi:Crosstalk} for B1 (in blue) and B2 (in red). The names and locations of the different IRs are also indicated.}
    \label{LHC_layout} 
\end{figure}

The first family is made of primary collimators (TCPs), which are the closest ones to the beams and have their jaws made of carbon-fiber-composite (CFC). The second family is composed of secondary collimators (TCSGs), also made of CFC, followed by the active absorbers (TCLAs), made of Inermet-180 (heavy Tungsten-alloy), which are placed to absorb particles out-scattered by the TCPs (secondary and tertiary beam halo). The last family is the tertiary collimators (TCTs) made of Inermet-180, which are installed upstream of the experimental insertion regions (IRs). These collimators aim to absorb the tertiary betatron beam halo and to provide passive protection of the aperture of the triplet quadrupoles of the final focusing system, as well as to control the experimental backgrounds. Downstream of the experiments other tertiary collimators are installed to absorb the debris from the collisions (TCLs). This three-stage hierarchy is installed in two insertions: insertion region 7 (IR7) for betatron cleaning and IR3 for off-momentum cleaning. Furthermore, two other CFC collimators per beam (TCSP, TCDQ) are installed in the extraction region (IR6) for beam dump protection. These collimators must ensure the protection of the machine in case of Beam Dump Failures (BDF)~\cite{assmann02dump,schmidt06,bruce15_PRSTAB_betaStar,bruce17_NIM_beta40cm}. The full 2018 LHC collimation layout is depicted in Fig.~\ref{LHC_layout} for the two counter-rotating LHC beams called Beam 1 (B1) in blue and Beam 2 (B2) in red.

The LHC operational cycle consists of different processes. The beam is injected into the ring with an energy of 450~$Z$~GeV. During the ``ramp and squeeze" the beam energy is increased up to its maximum of the run while the $\beta$-functions at the collision points (\bs) are decreased, which is called ``squeeze". Once at top energy (this static point in the cycle is referred to as ``flat top" in the following) the squeeze continues in a separate process to reach the minimum \bs, which is possibly different for the four experimental points.  Finally, in the last process that takes place after the end of the squeeze, the beams are brought in collision in the experiments in IR1/2/5/8 (see Fig.~\ref{LHC_layout}), the end point of this process is called ``physics''.

Before high-intensity beams are allowed in the machine, the performance of the collimation system is validated. This is done at each static point of the LHC cycle (injection, flat top, end of squeeze, and physics) by deliberately inducing losses using a safe, low-intensity beam and observing the resulting loss pattern.
For the betatron cleaning, losses are induced by blowing-up the beam in the transverse planes with the Transverse Damper (ADT) that can inject band-limited white noise in the beam~\cite{ipac11_hoefle_adt_blowup}. For the off-momentum cleaning validation, losses are induced by shifting the frequency of the radio frequency (RF) system. The losses occurring around the ring are recorded by beam loss monitors (BLMs)~\cite{holzer05,holzer08a}. Then, the BLM signals are plotted as a function of the location in the ring, $s$, and the losses are classified as cold (blue), warm (red) or collimator (black). The cold losses refer to losses in the aperture of SC magnets while the warm losses refer to losses in NC magnets and other equipment at room temperature. For the purpose of evaluating the collimation cleaning, the BLM signals are normalised by the highest BLM signal measured in the ring, which is typically measured in IR7 where primary beam losses are intercepted. The resulting loss distribution is called a loss map. These loss maps have been used to evaluate the agreement between measurements and simulations performed with the new \sixtrack-FLUKA coupling simulation tool. 

\begin{figure*}[!htp]
    \centering
    \subfloat[]{\includegraphics*[width=9.2cm]{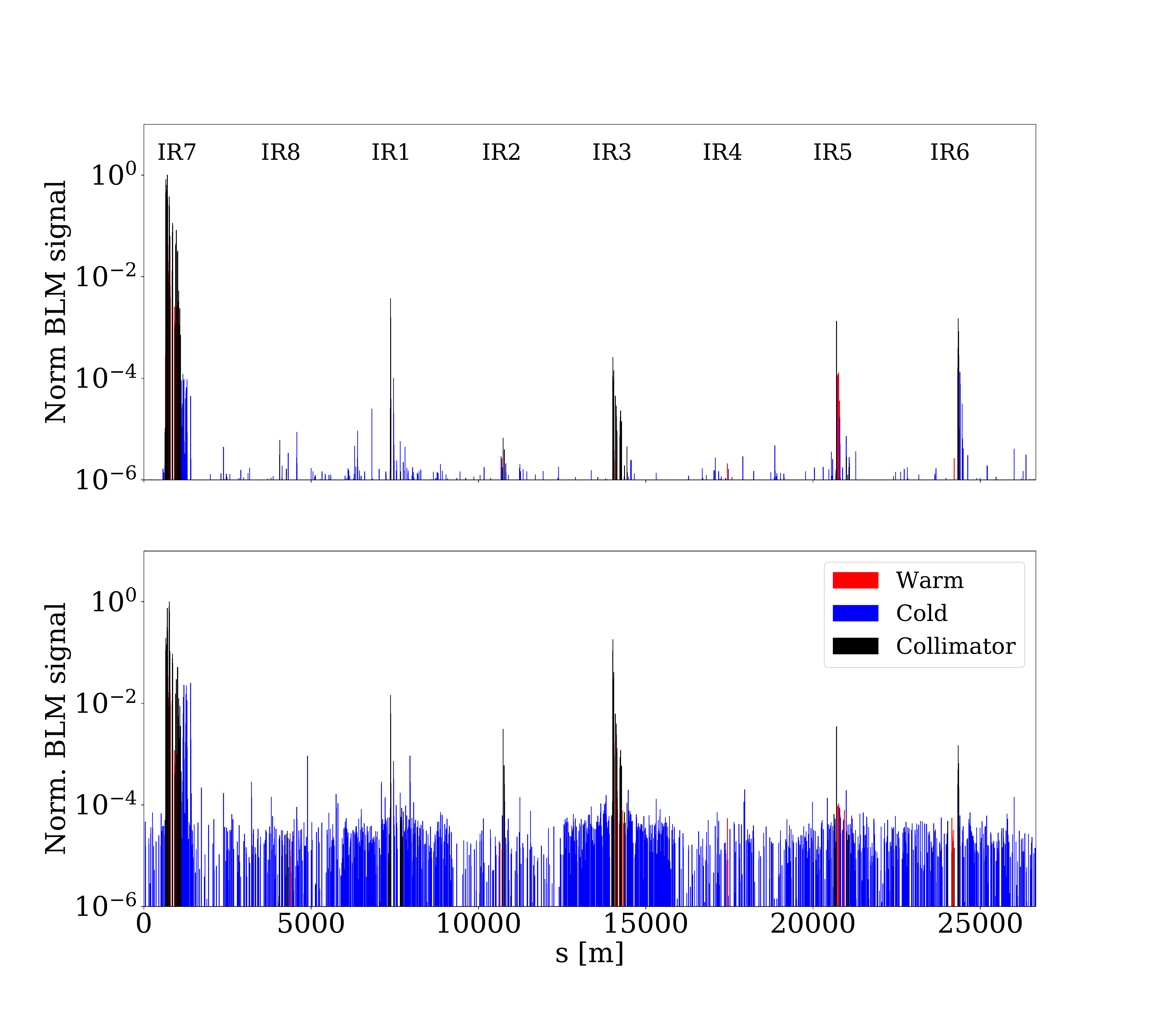} \label{protons_all}}
     \subfloat[]{\includegraphics*[width=9.2cm]{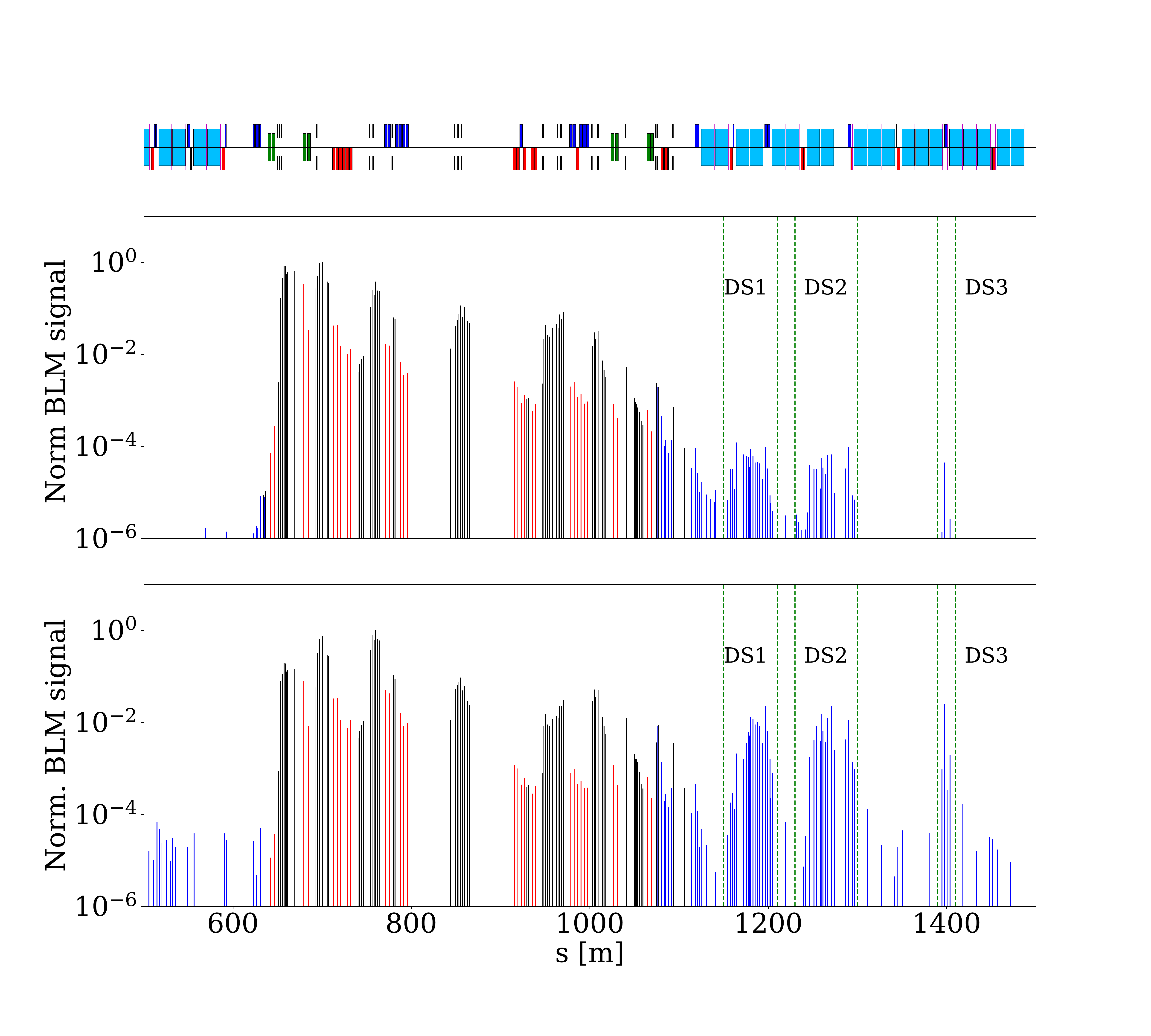}\label{ions_all}}
    \caption{Measured horizontal B1 full ring (a) and IR7 zoom (b) loss maps performed with colliding 6.5 TeV proton (top) and 6.37 Z TeV \Pb~ion (bottom) beams. In the IR7 zoom, the three clusters in the DS in which the highest cold losses spikes around the ring are observed, are indicated as DS1 ($s$=1150-1210 m), DS2 ($s$=1230-1300 m) and DS3 ($s$=1390-1410 m). Note that the layout has been started in IR7 with the vertical TCP located at $s$=650 m. On the top of (b) the different layout elements are indicated: dipoles (light blue), focusing quadrupoles (blue), defocusing quadrupoles (red), sextupoles (green) and collimators (black).}
    \label{LM_example} 
       \vspace*{-\baselineskip}
\end{figure*}

As an example, in Fig.~\ref{LM_example} the 2018 full ring (a) and IR7 (b) horizontal betatron loss maps for B1 are shown for protons (top) and \Pb~ions (bottom). These loss maps were performed with colliding beams optics. The highest cold spikes are found in three clusters downstream of the collimation system in the dispersion suppressors (DS) indicated in Fig.~\ref{LM_example} as DS1 ($s$=1150-1210~m), DS2 ($s$=1230-1300~m) and DS3 ($s$=1390-1410~m). As can be seen in Fig.~\ref{LM_example} the collimation cleaning efficiency in the three cold clusters in IR7 worsens by two orders of magnitude for \Pb~ions as compared to protons.

In the LHC, the validation of the collimation system performance is completed with asynchronous beam dump failure (ABDF) tests, which is a type of BDF scenario, in order to validate the protection of the machine by the collimation system during such failures. This critical failure occurs when there is a dump of the beam out of synchronisation with the abort gap (gap without beam that allows the extraction kicker magnets (MKDs) to rise up to full field). Mis-kicked bunches could cause fast high losses with consequent risk of damage of sensitive components. The most exposed elements are the collimators made of tungsten (TCTs and TCLAs) and the triplets that need to stay sufficiently behind the TCDQ and the TCSP collimators. These measurements are also an essential part of the beam commissioning after long periods without beam or following relevant changes in the hardware or in the machine configuration. For these tests, a single bunch is injected close to the abort gap and the orbit is bumped away from the TCDQ to simulate the maximum allowed orbit excursion in the extraction region. Then, the RF is switched off, thus allowing the beam to debunch and drift into the abort gap. This is monitored and when the beam fills the abort gap the beam is dumped by the operators. The resulting loss maps generated are analysed to evaluate the performance of the collimation system in such failure scenarios. This type of loss map has also been used to benchmark the simulation framework presented in this paper.

\begin{table*}[!htb]
\caption{\label{BeamSettings}B1 2015 and 2018 \Pb~main optics and beam run parameters at flat top and physics static points of the LHC cycle.}
\begin{ruledtabular}
\begin{tabular}{lcccc}
&Units& \textbf{\Pb }& \textbf{\Pb }&\textbf{\Pb }\\
&& \textbf{2018 Flat top} & \textbf{2018 Physics}&\textbf{2015 Quench test}\\
\hline
$\beta^*$ IR1/2/5/8& [m]& 1/1/1/1.5 & 0.5/0.5/0.5/1.5    & 1/1/1/3      \\
Half-crossing angle IR1/2/5/8& [$\mu$rad] & 160/200/160/-170   & 160/137/160/-170  & -145/137/145/-250     \\
Beam separation IR1/2/5/8& [mm]&1.1/6/1.1/2&0/0/0/0&1.1/4/1.1/2\\
IP shift IR1/2/5/8&[mm]&0/0/0/0 &0/-2/-1.8/0&0/0/0/0\\
E &[Z TeV]& 6.37  & 6.37     & 6.37     \\
$\epsilon_{norm}$ &[$\times$10$^{-6}$ m rad] &  1.39&  1.39    &  1.39    \\
\end{tabular}
\end{ruledtabular}
\end{table*}

\begin{table*}
\caption{\label{2018settings}2015 and 2018 \Pb~ion runs collimator settings for $\epsilon^P_N$=3.5~$\mu$m. L and R indicates the left and right jaw, respectively. H and V correspond to horizontal and vertical planes, respectively.}
\begin{ruledtabular}
\begin{tabular}{lccccc}
\textbf{Collimator} &\textbf{Beam} &\textbf{IR}  &  \textbf{\Pb~2018~Flat Top}    &\textbf{\Pb~2018~Physics}&\textbf{\Pb~2015 Quench test}\\
 & &  & ($\beta^*$ = 100 cm) &($\beta^*$ = 50 cm)&($\beta^*$ = 80 cm)\\
\hline
H TCP & B1     & 7  &5   &        5.5(L)-5.0(R) & 5.5\\
V TCP & B1     & 7  &5   &        5 & 5.5\\
H/V TCPs & B2     & 7  &5   &        5 & 5.5\\
TCSGs/TCLAs & B1/2     & 7  &6.5/10   &        6.5/10 & 8/14\\
TCP/TCSGs/TCLAs  & B1/2     & 3  &  15/18/20          &15/18/20&15/18/20  \\ 
H TCTs &B1      & 1/2/5    &15/15/15         &11/9/9 &37/37/37 \\ 
H TCTs &B2      & 1/2/5     &15/15/15          &9/9/9 &37/37/37 \\
V TCTs  &B1/2    & 1/2/5   &15/15/15            &9/9/9 &37/37/37 \\
TCTs    &B1/2   & 8       &15       &15 &37 \\
TCDQ / TCSP& B1      & 6    &7.4/7.4           &7.4/7.4 &9.1/9.1 \\ 
TCDQ / TCSP& B2        &6  &7.4/7.4          &7.4/7.4(L)-11.2 (R) &9.1/9.1 \\ 
TCL.4/5/6 &B1/2    & 1/5    &out/out/out         &15/15/out &out/out/out \\
\end{tabular}
\end{ruledtabular}
\end{table*}

Data from two different heavy-ion runs are used in this work.
The first set of data is from the most recent Pb-Pb run in 2018 at the energy of \qty{6.37\,Z}{TeV}~\cite{ipac19:IONRUN}. The second set of data used in the study is from a SC magnet quench test performed on the 13$\mathrm{th}$ of December 2015 with \Pb~ions~\cite{hermes16_ion_quench_test,Hermes:thesis}. This test was performed in order to measure the quench limit of the SC magnets in IR7. In such measurements, very high losses were produced at the primary collimator using the ADT, with the aim of quenching the IR7 DS magnets with the collimator debris in a controlled manner. B2 and the horizontal plane were used for the experiment in which, for the first time in the LHC, a magnet quench was achieved with beam losses from collimators in a controlled way. The better understanding of the quench limit and agreement with expectations allows pushing the maximum stored beam energy in the machine and defines upgrade requirements for the LHC collimation system. These measurements offered a unique opportunity to benchmark simulations and measurements with very high resolution as the high losses cause a high signal-to-noise ratio. The main beam parameters and collimator settings for both runs are summarised in Tables~\ref{BeamSettings} and~\ref{2018settings}.

To calculate the collimator settings in units of beam size, $\sigma$, the normalised LHC design emittance for protons, $\epsilon^P_N$=3.5$\times$10$^{-6}$ m~rad is used. In order to define comparable collimator settings for the ion beams the same geometrical emittance has to be used. Therefore an equivalent normalised ion emittance $\epsilon^I_N$ is calculated taking into account their corresponding relativistic factor, $\gamma_I$, such that the following relation is satisfied:
\begin{equation}
    \frac{\epsilon^P_N}{\gamma_P}=\frac{\epsilon^I_N}{\gamma_I}
    \label{emit}
\end{equation}
The normalised emittance value calculated using Eq.~\eqref{emit} for \Pb~is shown in Table~\ref{BeamSettings} and used to generate the corresponding beams for the simulations presented in Section~\ref{sec:benchmark}.

\section{\label{sec:simulation_method}Heavy-ion collimation simulations}

Heavy-ion collimation simulations must include both a high-precision magnetic tracking as well as a good modelling of the different nuclear interaction processes of heavy ions with the collimator material, in particular EMD and nuclear fragmentation, as well as an accurate tracking of the nuclear beams and out-scattered, off-rigidity fragments through the magnetic lattice.

A first software tool, ICOSIM~\cite{ICOSIM} developed in 2004, allowed us to perform the magnetic tracking in linear approximation using lookup tables of cross sections for the creation of fragments in collimators. It tracks only the heaviest ion fragment created in each interaction. The comparison of loss maps simulated with ICOSIM and the LHC BLM measurements showed that the physics models and approximated optics used in this software are adequate to identify some collimation issues, but not detailed enough to model accurately the collimation of heavy ions~\cite{Hermes:thesis,hermes_nim}. A later tool, STIER (\sixtrack with Ion-Equivalent Rigidities)~\cite{Hermes:thesis,hermes_nim} improved the accuracy of the tracking and the ion-matter interactions, but it modelled only the first collimator hit in detail. All subsequent collimators were modelled as perfect absorbers, and in the tracking the ions were modelled as protons with equivalent magnetic rigidity.

In this article, we present a new simulation tool that on one hand can accurately track any heavy ion species without the proton approximation, and on the other hand simulates the ion-matter interactions for all tracked particles in all collimators. To achieve this, the magnetic tracking is done by \sixtrack, which provides a 6-dimensional tracking of relativistic beams in high-energy synchrotrons over many turns based on symplectic tracking maps. A more generic approach of computing the heavy-ion trajectories was developed by including in \sixtrack new maps for the tracking of heavy ions, as described in Appendix A. 

The ion-matter interactions are handled by the Monte Carlo code FLUKA~\cite{fluka-web} that has the most up-to-date heavy-ion physics~\cite{PhysRevSTAB.17.021006}. FLUKA is used to simulate particle transport and the particle-matter interactions in a user-defined 3D geometry. Full online coupling between \sixtrack and FLUKA was used~\cite{mereghetti13_ipac,skordis18_tracking_workshop}. This is described further in Section~\ref{sec:coupling}. 

\subsection{Coupling between SixTrack and FLUKA}
\label{sec:coupling}
\begin{figure}[!htb]
    \centering
    \includegraphics*[width=\columnwidth]{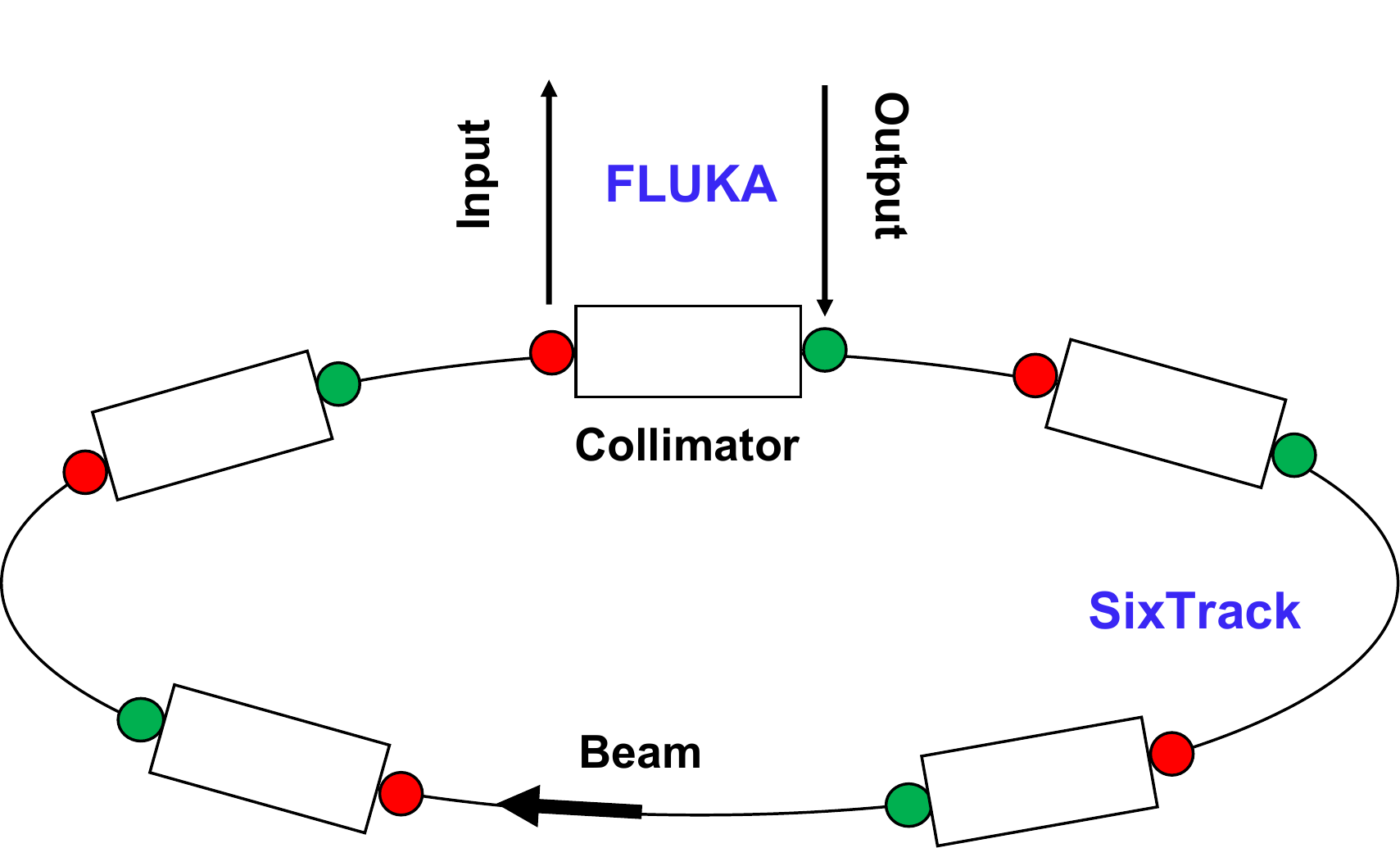}
    \caption{\sixtrack-FLUKA coupling principle.}
    \label{SFC_scheme} 
\end{figure}

The existing framework for coupling \sixtrack and FLUKA~\cite{mereghetti13_ipac,skordis18_tracking_workshop} for protons was expanded to incorporate multi-isotopic tracking~\cite{Hermes:thesis}. The two simulation codes are run in parallel and the exchange of particles between them is done through a network port and a C library to handle the communication between the codes. This significantly shortens the required simulation time compared to a reinitialisation of each code after each particle exchange. The basic principle is illustrated in Fig.~\ref{SFC_scheme}. At the start of every collimator there is an extraction marker (red dots in Fig.~\ref{SFC_scheme}), and at this position the particle co-ordinates are sent to FLUKA where the interaction with the collimator is simulated. The implementation is done in such a way that the number of particles sent back to SixTrack from FLUKA can be both larger or smaller than the number of incoming particles, in order to cover for cases where ion fragments (or other secondary particles, if requested) are created or absorbed. When all input ions have been tracked through the collimator, the output distribution of particles at the end of the collimator is sent back to \sixtrack and re-injected into the lattice at the dedicated injection markers (green dots in Fig.~\ref{SFC_scheme}) from which the tracking is continued in \sixtrack. The markers are set at the beginning and at the end of each collimator tank. 

The user set-up for the FLUKA simulation part is done through the generation of a standard FLUKA input file where the 3D geometry of each collimator or user-defined insertion device is defined. In this file the user can adjust the physics settings for the simulations, such as energy cuts and transport thresholds to be applied, or switch off unnecessary physical processes to optimise the computational time. The particle types sent back from FLUKA to \sixtrack are also defined by the user. Typically, for heavy-ion collimation studies only nuclear fragments are considered because the fraction of energy carried by other particles is very small (see for more detail Section~\ref{sec:level3}). 

To record losses on other elements that are not modelled in FLUKA, such as magnets, an online aperture check has been developed~\cite{mereghetti13_ipac,skordis18_tracking_workshop}, which alongside the tracking checks the trajectories of all tracked particles against an aperture model with a default precision value of 10~cm, that can be set to any other desired value by the user. The tracking stops as soon as the aperture is crossed and the impact point is recorded as well as the ion type and energy carried.

\section{\label{sec:level3}Simulations setup and analysis}

We discuss two different types of simulations in this article: betatron beam halo collimation cleaning, and BDFs scenarios.

For halo cleaning studies, the number of simulated halo particles is about 3--6 $\times$10$^6$ initial heavy-ions depending on the case. Our simulations have been performed with a monochromatic beam since the energy spread is negligible for the loss pattern generated after the interaction of the main beam with the primary collimator. In the non-collimation transverse plane a Gaussian distribution in the range 0--3$\sigma$ is generated, while in the collimation plane a beam halo distribution matching the phase space ellipse at the primary collimator is generated following the method outlined in~\cite{LHCcoll:5}. In addition, only the heavy-ions with a given impact parameter, $b$, defined as the distance between the heavy-ions impacting the collimator and the surface of the collimator are selected. This means that the halo particles hit the collimator already on the first turn, and the physical mechanisms for the slow diffusion to larger amplitudes (e.g. electron cloud, beam-beam interactions, and intra-beam scattering to name a few) are not simulated, as this would require too much computing power and it is not relevant to evaluate the collimation system cleaning efficiency. 

\begin{figure}[!htb]
    \centering
    \includegraphics*[width=\columnwidth]{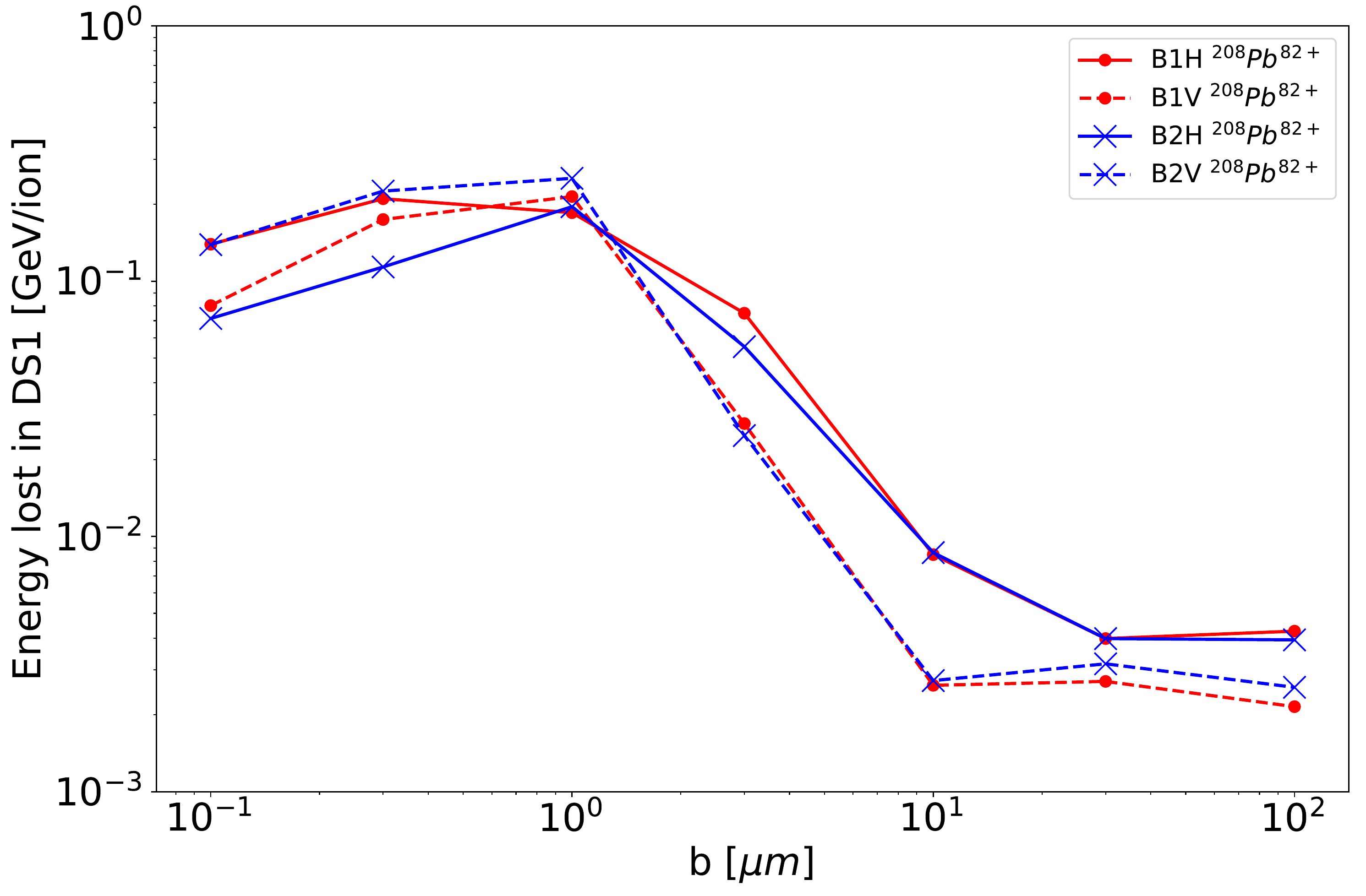}
    \caption{Energy lost in the DS1 cluster as a function of $b$ for 6.37~$Z$~TeV horizontal and vertical \Pb~ion beams using the 2018 colliding optics.}
    \label{bscan} 
\end{figure}

The actual value of $b$ in the LHC depends on the beam loss process and it is not well known and may even vary between runs. Simulations were performed with different impact parameters for the colliding 2018 \Pb~optics for both beams and both planes. 
Fig.~\ref{bscan} shows the energy lost in the IR7 DS1 (defined between s=1150-1210 m in Fig.~\ref{LM_example}) as a function of $b$. 
From this study we could conclude that for $b$ = \qty{1}{\mu m}, the energy lost in DS1 is maximised for all cases studied. Because of that, all simulations presented in the following have been performed with $b$ = \qty{1}{\mu m}, in order to stay on the pessimistic side. This is therefore a conservative approach for this study and for predictions of performance for future configurations. 

\begin{figure*}[!htb]
    \centering
     \subfloat[]{\includegraphics*[width=9cm]{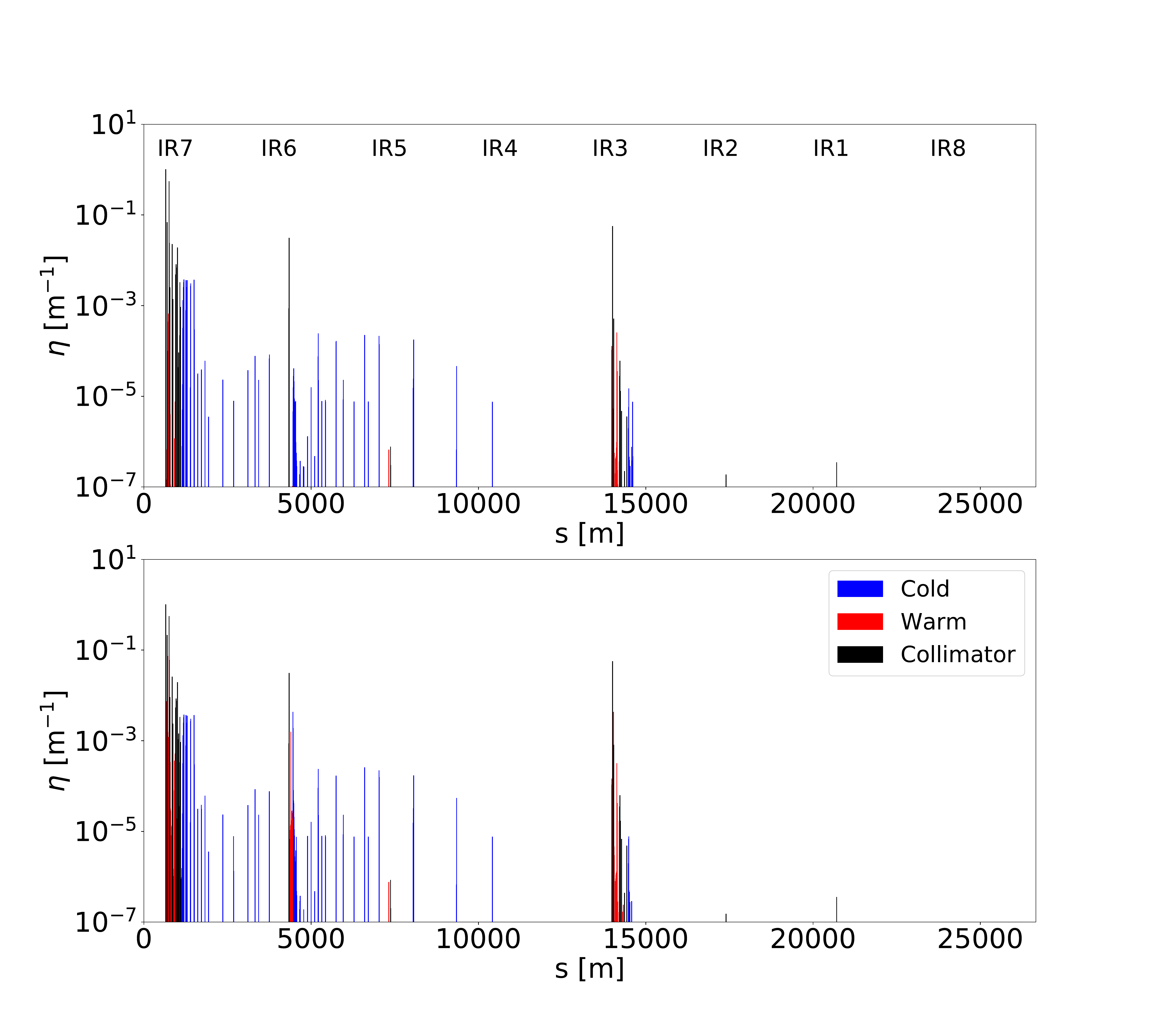} \label{protons_all}}
      \subfloat[]{\includegraphics*[width=9cm]{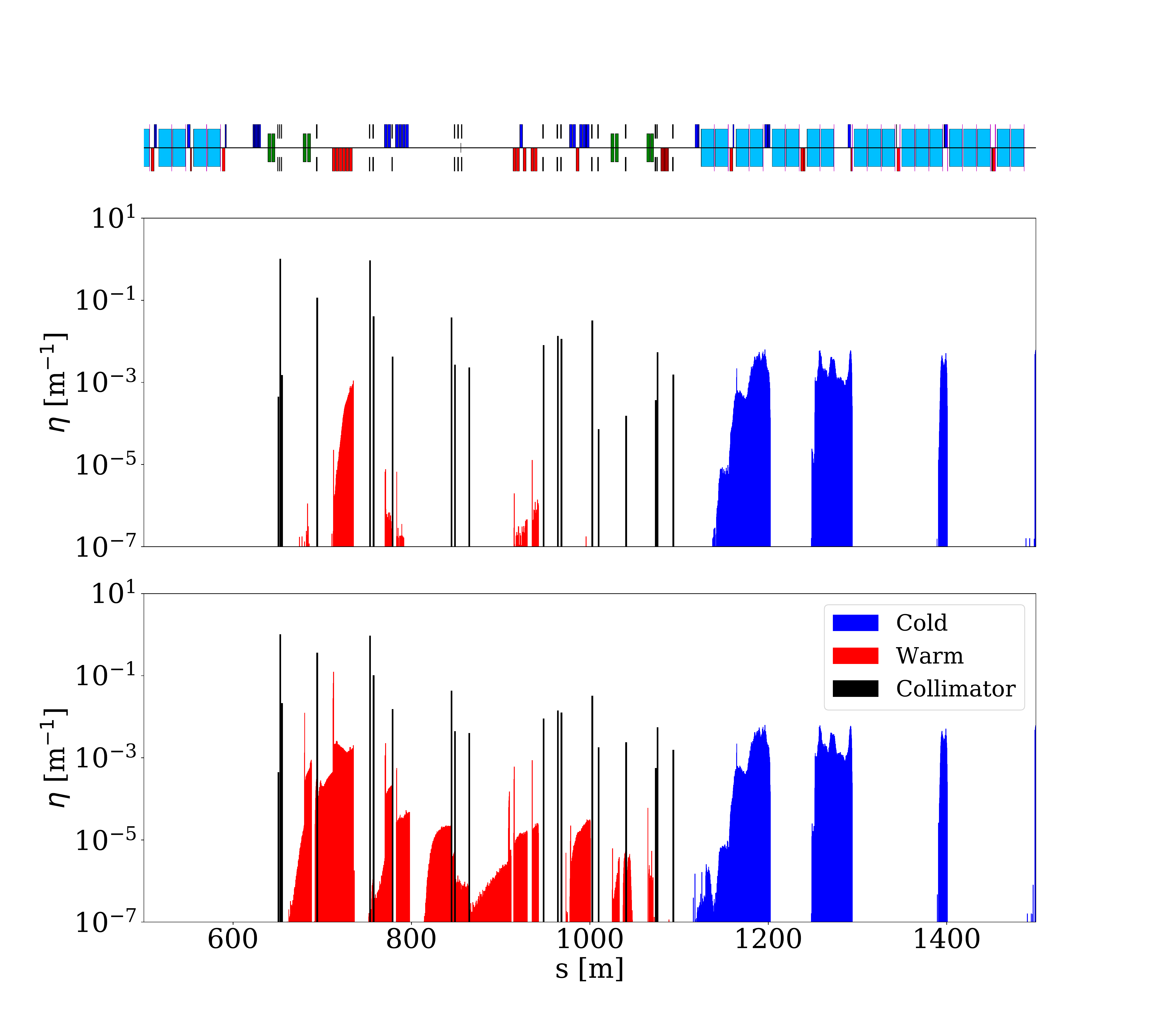} \label{protons_zoom}}
    \caption{B2 horizontal full ring (left) and IR7 zoom (right) loss map for 2018 \Pb~ion colliding optics without (top) and with (bottom) the tracking of the protons generated at collimators.}
    \label{protons_example} 
\end{figure*}

For the purpose of the cleaning studies performed with the \sixtrack-FLUKA coupling, which are different to local energy deposition studies, no magnetic rigidity cut is applied on the produced fragments, while a particle transport threshold cut of 1 TeV per nucleon is considered. The particles with an energy lower than the 1 TeV per nucleon threshold are assumed to be lost at the collimator. This allows us to speed up the simulation without altering the result, since fragments affected by the cut will be lost very close to the production point anyway. Physics processes for nuclear interactions, ionisation, EMD, nuclear evaporation, and statistical fragmentation are activated, while the electromagnetic shower generation is not considered. Note that the electromagnetic shower will only affect the local energy deposition studies that could be further studied in dedicated FLUKA simulations. In Section~\ref{sec:2015_quench_test}, an example of detailed IR7 FLUKA simulations including the electromagnetic showers is presented and compared with the results from the coupling tracking tool. 

The energy lost at the collimators can be calculated as the difference between the energy of all incoming particles, passed to FLUKA for the interaction, and the surviving particles sent back to the tracking. However, in this case the collimator losses are overestimated, since some energy is carried by particles that exit the collimator, but are not sent back to \sixtrack. In order to correctly estimate the energy deposited in the collimators, the energy carried by these particles is saved in a file at the exit marker of the collimator. Then, this information is used in the post-processing to correct the losses at the collimators. The two major contributing particle types are neutrons and protons. 

To illustrate this and understand the impact of including or not the tracking of protons generated at the collimators, simulations have been performed with an initial population of 3$\times$10$^{6}$ \Pb~ions and the 2018 collision optics. 

From the simulation results, giving the loss positions around the ring, the loss maps are constructed calculating the cleaning inefficiency, $\eta(s)$, as the sum of the energy lost at a given location, $s$, and per unit length normalised by the maximum energy lost in the ring within a distance $\Delta s$, $E_\mathrm{max}$, as
\begin{equation}
    \eta(s)=\frac{\sum_i E_i(s)}{\Delta s~E_{max}},
\end{equation}
where $E_i$ is the energy of the ion $i$ lost within a distance $\Delta s$ around the position $s$. For losses in warm and cold elements, we use $\Delta s=10$~cm, corresponding to the chosen resolution of the online aperture check algorithm, while for collimators it corresponds to the collimator active length of 0.6 m for TCPs, 1 m for TCSPs, TCLAs, TCTs, TCLs and TCSPs; and 9 m for the TCDQ.

In Fig.~\ref{protons_example}, the full ring (left) and IR7 zoom (right) horizontal loss maps for B2 2018 \Pb~collision optics are shown without (top) and with (bottom) the tracking of the protons. Small variations in the losses at the first secondary collimators in IR7 (see Fig.~\ref{protons_zoom}) are observed as well as more warm losses (red) in IR7, IR6 and IR3 due to the generation of protons in the collimators. In the IR7 TCP made of CFC about 14 protons are produced per impacting ion. These protons are lost close to the origin collimator in warm elements as can be seen in IR7 due to the magnetic rigidity difference with respect to the reference heavy-ion of more than 99$\%$. The same happens to the protons originated at the TCSP and at the TCP in IR6 and IR3, respectively. The magnitude of the observed differences in collimators and cold losses is not significant when the simulation and beam measurements are benchmarked, as shown in Section~\ref{sec:benchmark}. Note that despite the correction for proton and neutron energies, the result is still not the real energy deposited in the collimator since the energy can escape in the form of electromagnetic and hadronic showers (e.g. pions, kaons, electrons, etc), which are not simulated.
In the following simulations presented in this paper, particles lighter than deuterium are not sent back to \sixtrack as it is expected that losses are mainly local in IR7 while the simulation and post-processing time is considerably increased. 

The \sixtrack-FLUKA coupling framework can also be used to perform simulations of BDF scenarios. For these simulations a special \sixtrack module~\cite{sjobak15_ipac_dynk, sjobak_trackingWS_dynk} is used to change dynamically the field of the MKDs during the simulation, using realistic time-dependent kicks based on measured wave forms~\cite{fraser16_ipac}. As in Ref.~\cite{bruce17_NIM_beta40cm}, where similar studies were performed for proton beams, we simulate several consecutive Gaussian bunches centred on the nominal closed orbit, each encountering different MKD strengths. For the results presented in this paper a 75 ns bunch spacing is considered, this implies that 44 bunches are kicked during the rise of the MKD fields. The first bunches receive small kicks and pass through the whole ring, later bunches receive large kicks and hit the dump protection collimator or are extracted from the machine while the intermediate bunches are the ones that risk hitting the machine aperture and sensitive collimators. Because of that, only a small range of intermediate bunches (between 5-25) are simulated using a separate three-turn simulation for each bunch. In the first turn, no MKD kick is implemented. In the second turn the simulated bunch is affected by the intermediate MKD kicks, different for each bunch. In the third turn the maximum MKD kick value is applied and the particles remaining on this turn are lost in the extraction point. Note that only the ring losses are of interest in this study and not the losses in the extraction line. We present the final loss distributions as the sum over all simulated bunches, normalised to the absolute energy of the lost particles and scaled to the real intensity in the machine for the number of bunches simulated. 
 
Different BDF modes can occur~\cite{Assmann:691845}: the single module pre-fire (SMPF), in which one MKD module spontaneously fires first and the remaining MKDs are then automatically triggered after a short delay, and the ASBD mode in which all MKDs fire simultaneously but at the wrong time when the beam is passing. This last mode is the one measured in the machine and for which comparison studies with simulations have been performed and presented in Section~\ref{sec:benchmark}.

\section{Applications and comparison of simulations with measurements}
\label{sec:benchmark}

\begin{figure*}[!htb]
    \centering
    \includegraphics*[width=17cm]{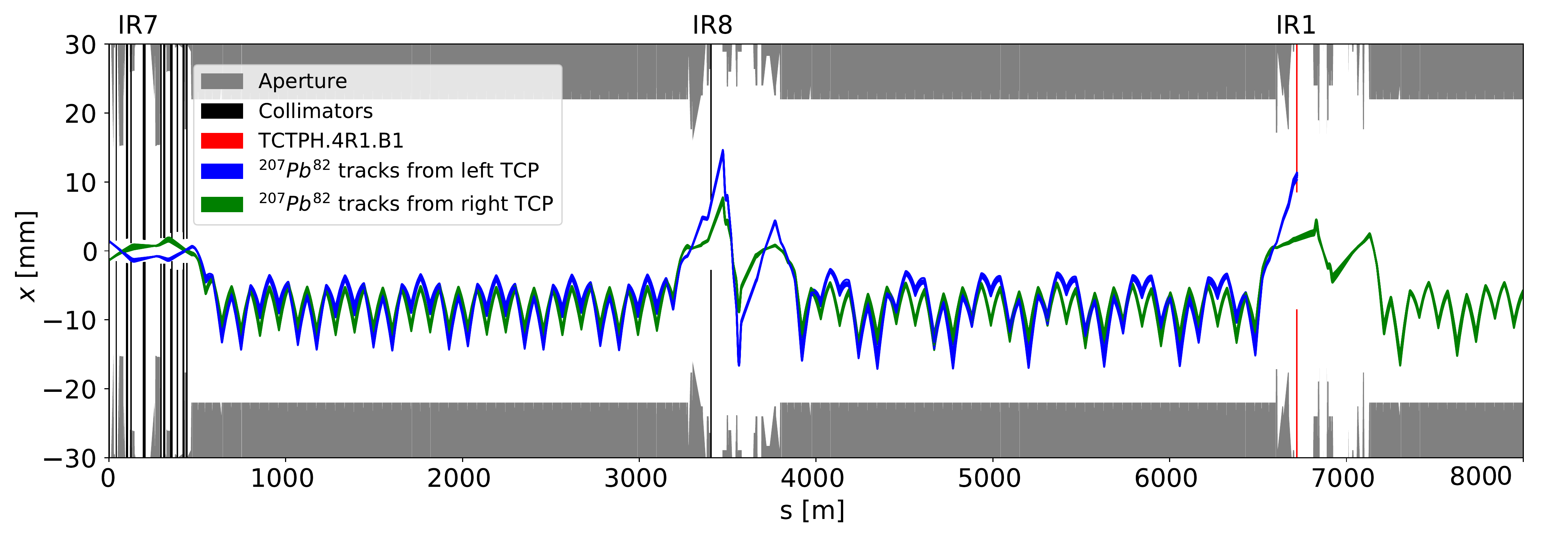}
    \caption{\isotope{207}{Pb}{82+}~ion orbit from the TCP to IR1 together with the aperture model and collimators with TCTPH.4L1.B1 at 9$\sigma$ half gap highlighted in red for the 2018 \Pb~ion run physics optics.}
    \label{Pbtrack} 
\end{figure*}
\begin{figure*}[!htb]
    \centering
    \subfloat[]{\includegraphics*[width=6cm]{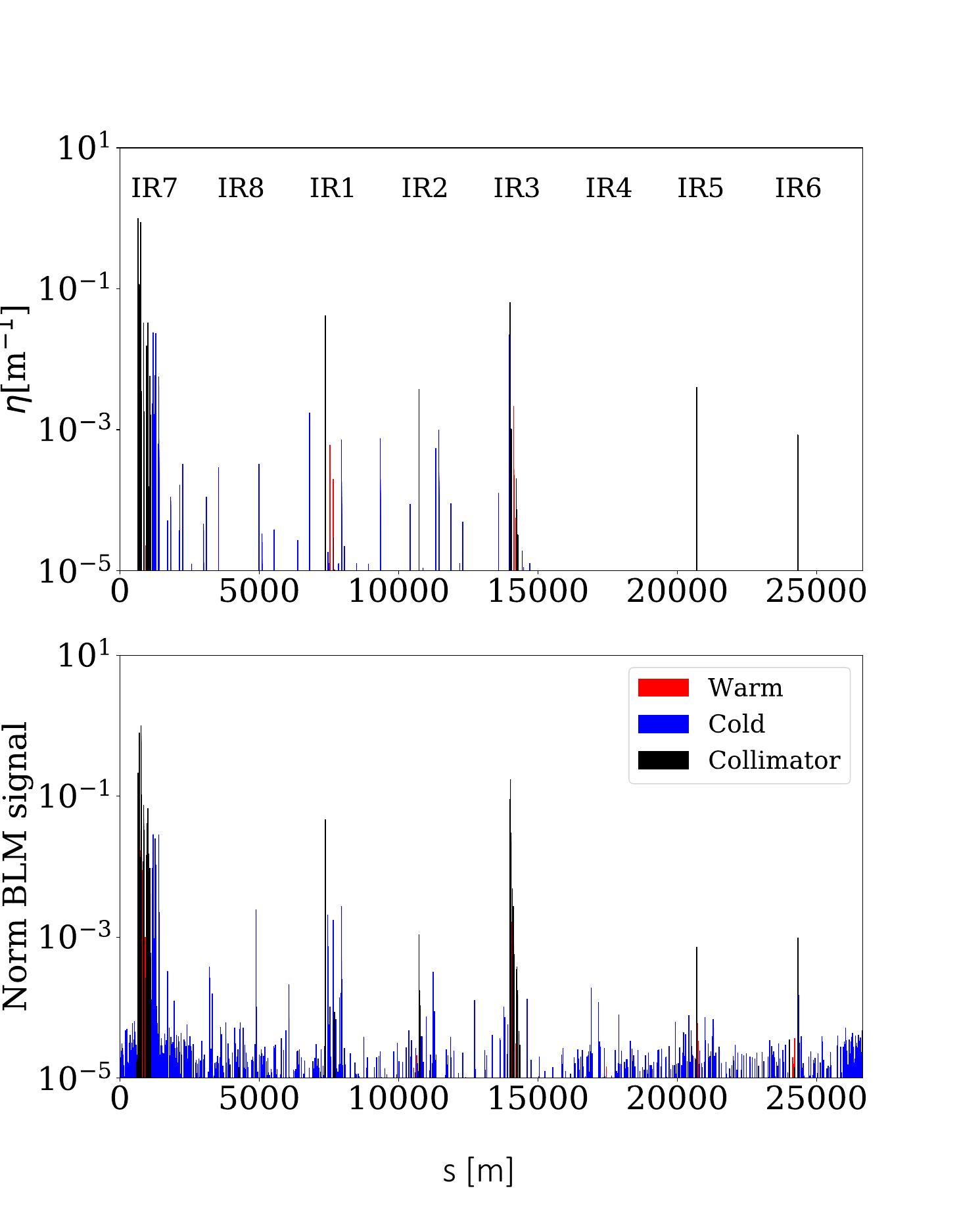} \label{Pb_both}}
    \subfloat[]{\includegraphics*[width=6cm]{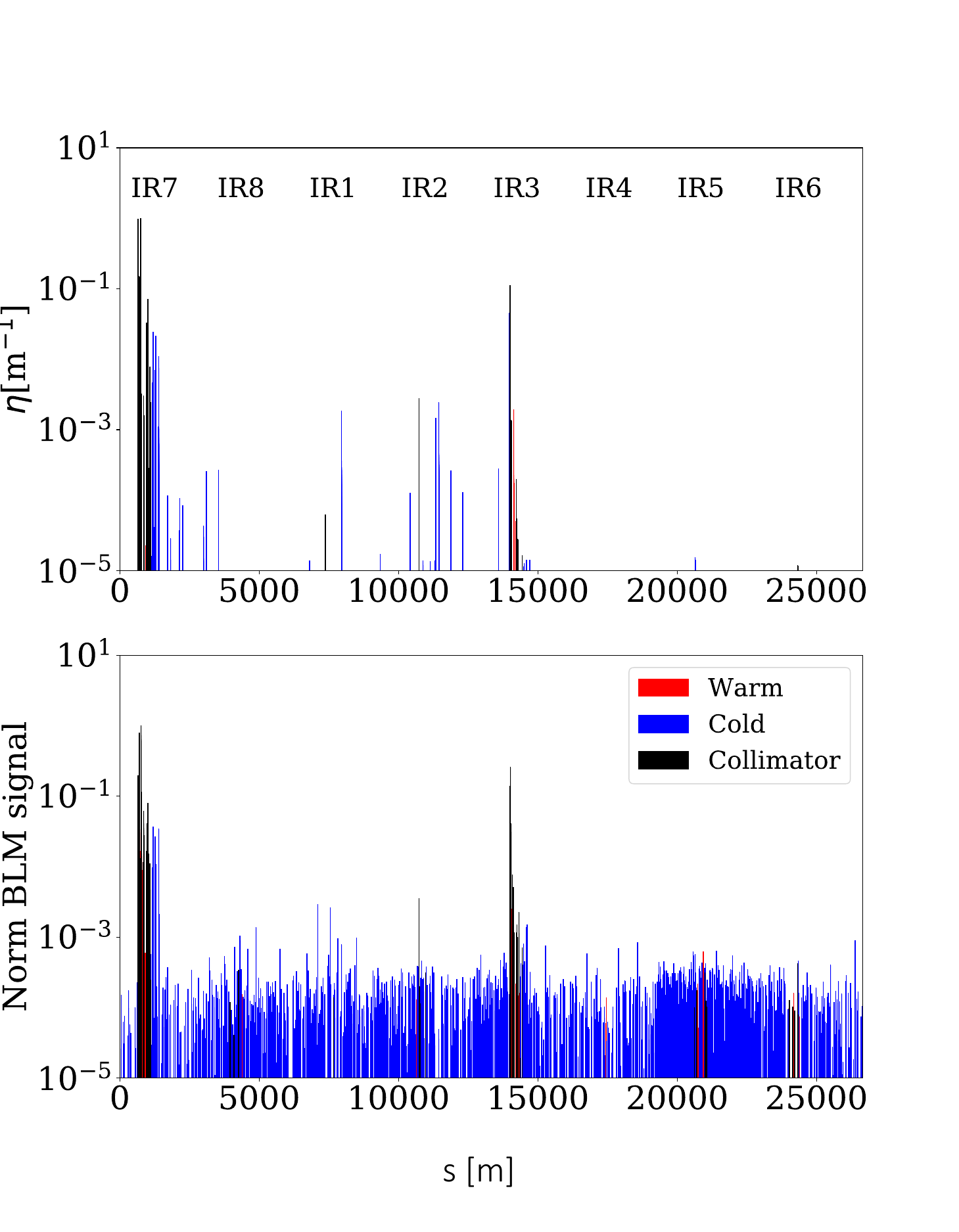}\label{Pb_R}}
     \subfloat[]{\includegraphics*[width=6cm]{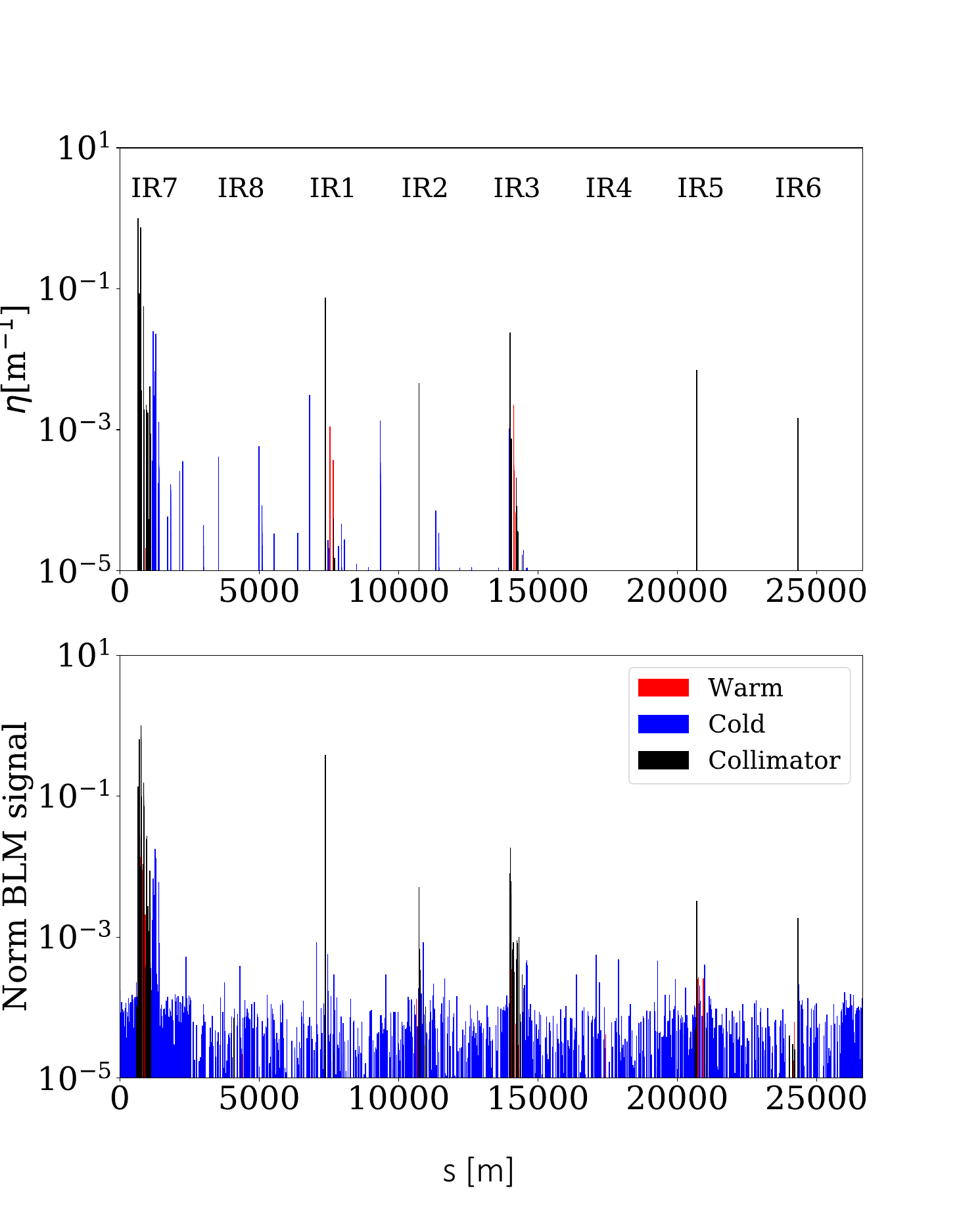}\label{Pb_L}}
    \caption{Horizontal B1 simulated (top) and measured (bottom) loss map with both TCP jaws closed to 5$\sigma$ (a), with only the right TCP jaw closed to 5$\sigma$ (b) and with only the left TCP jaw closed to 5$\sigma$ (c).}
    \label{Pb_asym} 
\end{figure*}

In this section, we show examples of applications of our simulation setup to the LHC operation during the 2015 and 2018 Pb-Pb runs. The simulations demonstrated to be a very good guide to understand the origin of the losses in various complex configurations and to optimise the collimator settings and formulate mitigation strategies. Simulated and measured loss maps for the betatron cleaning and ABDF scenarios are compared and the agreement is discussed. It should be noted that quantitative discrepancies up to at least a factor 10 can be observed in the comparison of simulated and measured loss maps~\cite{LHCcoll:5}. This is because the simulated loss maps show the sum of the energy of the lost particles impacting on the aperture around the ring, while the measured losses are taken from the BLM signals. Since the BLMs measure the secondary particles created in the showers, the ratio between local losses and BLM signal can vary significantly between different locations, depending on the materials and local geometry as well as the impact distribution.

A more quantitative comparison between simulation and measurements can be obtained with energy deposition simulations of the local showers at critical loss locations (e.g. with FLUKA), including the local geometry and the BLM and hence better representing the BLM response. However, this is only possible for a few locations and not for the full ring due to the required resources to prepare detailed local geometries and to the computational time. In order to illustrate the improved quantitative agreement when performing detailed FLUKA simulations, IR7 results obtained with the full simulation chain (i.e. tracking and showering) for the 2015 Pb-Pb collimation quench test are presented in Section~\ref{sec:2015_quench_test}. For the purpose of this paper, it will be shown that the comparison based on loss maps is adequate.

\subsection{ 2018 \Pb~ion run}
During the commissioning of the 2018 Pb-Pb run, losses were observed at the horizontal TCT (TCTPH) in IR1 in the physics configuration (see Table~\ref{2018settings}) that could potentially cause high background or beam dumps~\cite{PbIPAC_Nuria}. An effort was made to decrease these losses and the \sixtrack-FLUKA simulations were then used to understand the loss sources and to propose mitigation measures.

The simulations showed that the losses at the TCTPH consisted mainly of \isotope{207}{Pb}{82+}~ions, which were created inside the left TCP jaw (while moving in the beam direction) and then bypassed all other collimators. The \isotope{207}{Pb}{82+}~ion is the fragment with the smallest rigidity offset and it stays within the acceptance of the arcs, so it can travel far along the machine circumference. In Fig.~\ref{Pbtrack} the tracks of  \isotope{207}{Pb}{82+}~ions originated on the left (blue) and right (green) TCP jaw are shown from the TCP to IR1 together with the aperture model and collimators with the TCTPH indicated in red. Fragments emerging from the two TCP jaws are lost at different locations, because the betatron motion is either amplifying or compensating the dispersive offsets depending on the starting betatron phase. This was an important result from simulations that was tested and applied in LHC operation. Note that this could not be found out only through measurements as it is not possible to test many different collimator configurations during the commissioning phase due to time constraints. Therefore, accurate simulation results are crucial to anticipate possible issues and elaborate loss mitigation strategies. 

We therefore proposed to decrease the TCTPH losses by either slightly retracting the TCTPH jaws, which was possible since there was significant margin to the triplet aperture, or by concentrating the primary losses on the right TCP jaw only. The latter option could be achieved by retracting the left TCP jaw by a small amount. Figure~\ref{Pb_asym} shows a comparison of the simulated (top) and measured (bottom) horizontal loss maps for B1 with both TCP jaws closed to 5$\sigma$ (Fig.~\ref{Pb_both}), with only the right TCP jaw closed to 5$\sigma$ and the left jaw opened to 36$\sigma$ (Fig.~\ref{Pb_R}) and with only the left TCP jaw closed to 5$\sigma$ and the right jaw opened to 36$\sigma$ (Fig.~\ref{Pb_L}). 

A very good qualitative agreement between simulations and measurements is observed on the losses at the collimators. The losses at the TCTPH in IR1 (at $s\approx$ 7500~m) are reduced by more than three orders of magnitude when the left TCP jaw is opened in both simulated and measured loss maps. In addition, when the left TCP jaw is opened, no losses are observed at the TCTPH in IR5 and at the TCSP in IR6 while the losses in the TCP in IR3 slightly increase, in both simulations and measurements. Quantitatively, discrepancies up to one order of magnitude are observed, for example on the level of losses in the TCTPH in IR1 when only the left TCP jaw is considered. These levels of discrepancies are in the order of what is expected due to the difference in BLM response as discussed in the introduction of this section. In addition, note that no orbit and machine errors are taken into account in the simulations which could also contribute to these observations. 

\begin{figure}[!htb]
    \centering
    \includegraphics*[width=9cm]{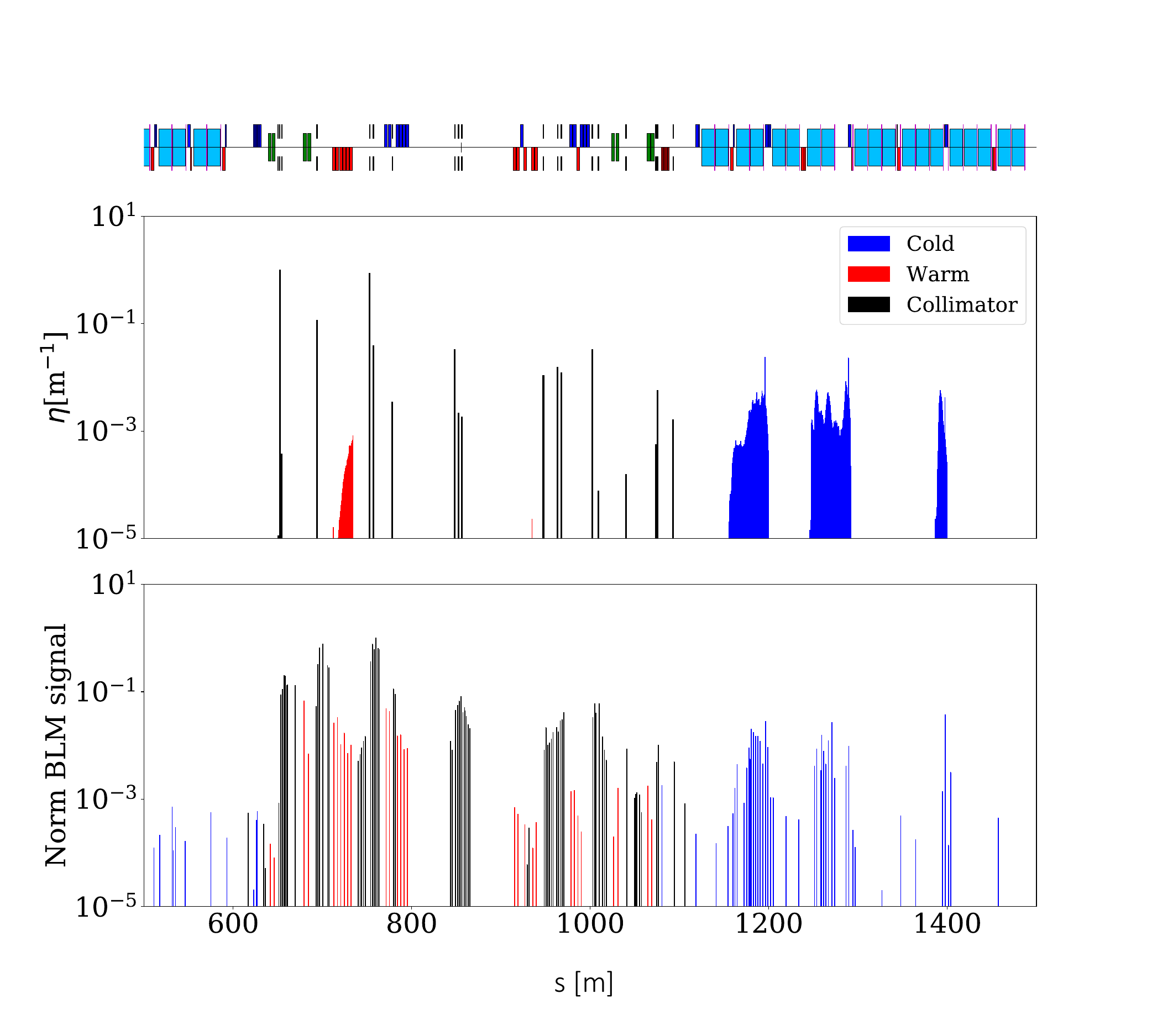}
    \caption{IR7 horizontal B1 simulated (top) and measured (bottom) loss map for both TCP jaws at 5$\sigma$.}
    \label{Pb_IR7} 
\end{figure}

Concerning cold losses, spikes between IR7 and IR1 when both TCP jaws are in place, as seen in Fig.~\ref{Pb_both}, are observed in both simulations and measurements. These losses are mainly coming from the left TCP jaw. The longitudinal location of these spikes is in some cases shifted or even absent in one of the two (measured or simulated). These differences could be explained by aperture or orbit displacements in the real machine that are not reproduced in simulations. However, the location with the highest amount of cold losses is very well reproduced as can be seen in Fig.~\ref{Pb_IR7}, where the simulated (top) and measured (bottom) IR7 loss maps are shown. The losses at the three clusters, named in Section~\ref{sec:level2} as DS1, DS2, and DS3, where the locally generated dispersion function from the TCP is beating with amplitudes between 0--2.1~m, are very well reproduced by the simulations. Concerning the losses at the collimators in IR7, quite good agreement can be observed on the hierarchy, however in measurements the signals are higher at the secondary collimators than at the primary collimators, but not in simulations. This difference with respect to the simulations could be due the fact that IR7 BLMs intercept showers also from losses further upstream and to the BLM response, accounted for by a second step of additional FLUKA simulations discussed in Section~\ref{sec:2015_quench_test} where the quantitative agreement is significantly improved.

\begin{figure}[!htb]
    \centering
    \includegraphics*[width=\columnwidth]{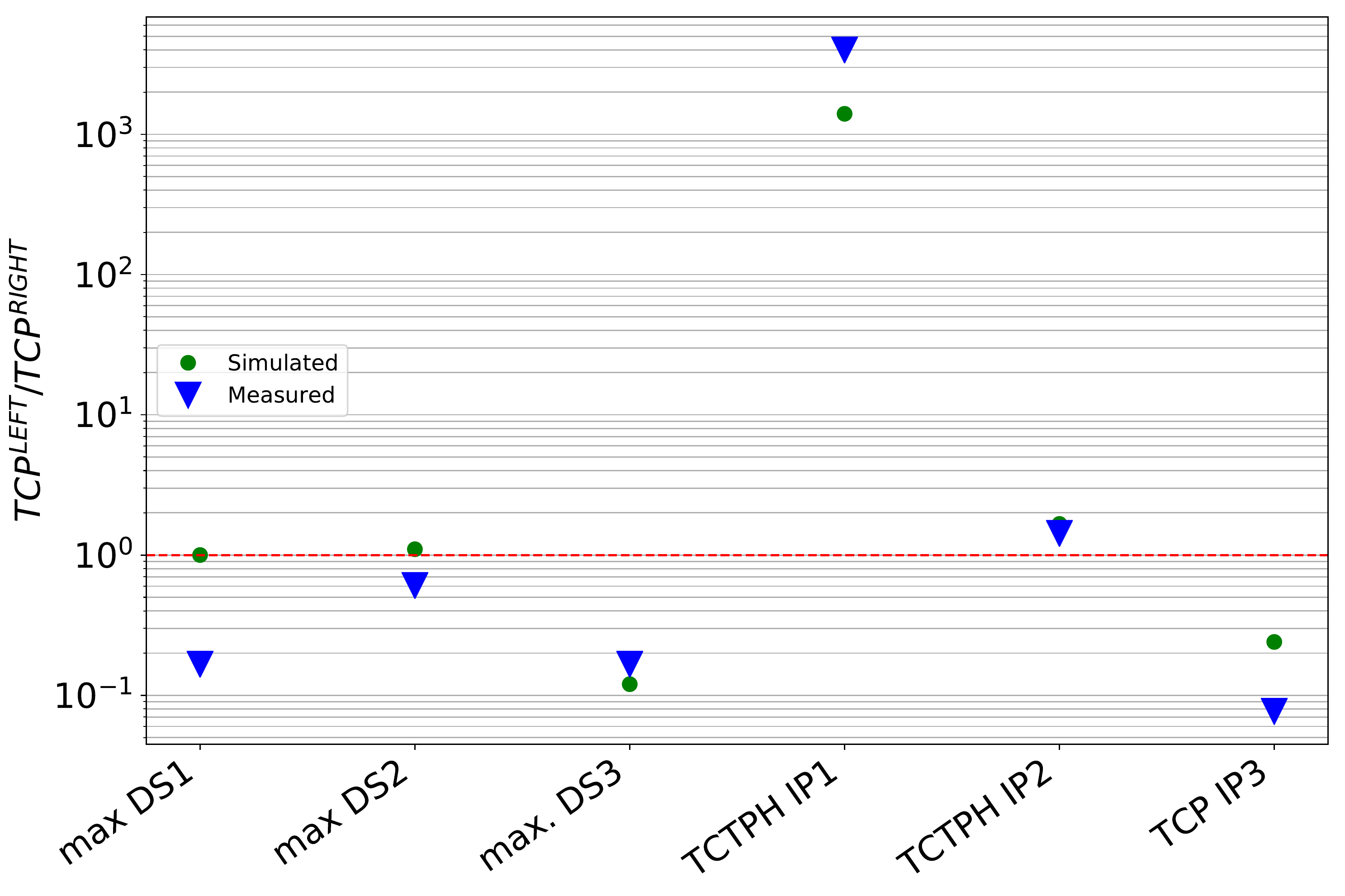}
    \caption{Simulated (green) and measured (blue) ratio of losses generated by the left and right TCP jaws at 5$\sigma$ at different locations. Note that only
collimators with BLM signals above the noise level are considered.}
    \label{Pb_asym_summary} 
\end{figure}

To illustrate the differences between the loss map generated by only the left TCP jaw and only the right TCP jaw at 5$\sigma$, Fig.~\ref{Pb_asym_summary} shows the ratio of losses obtained in simulations (green) and measurements (blue) between the two loss maps at different locations. In general, a very good agreement is observed with the trend in all depicted locations between simulations and measurements, in particular since the BLM response is not included. Note that the data cover several orders of magnitude.

\begin{figure}[!htb]
    \centering
    \includegraphics*[width=9.4cm]{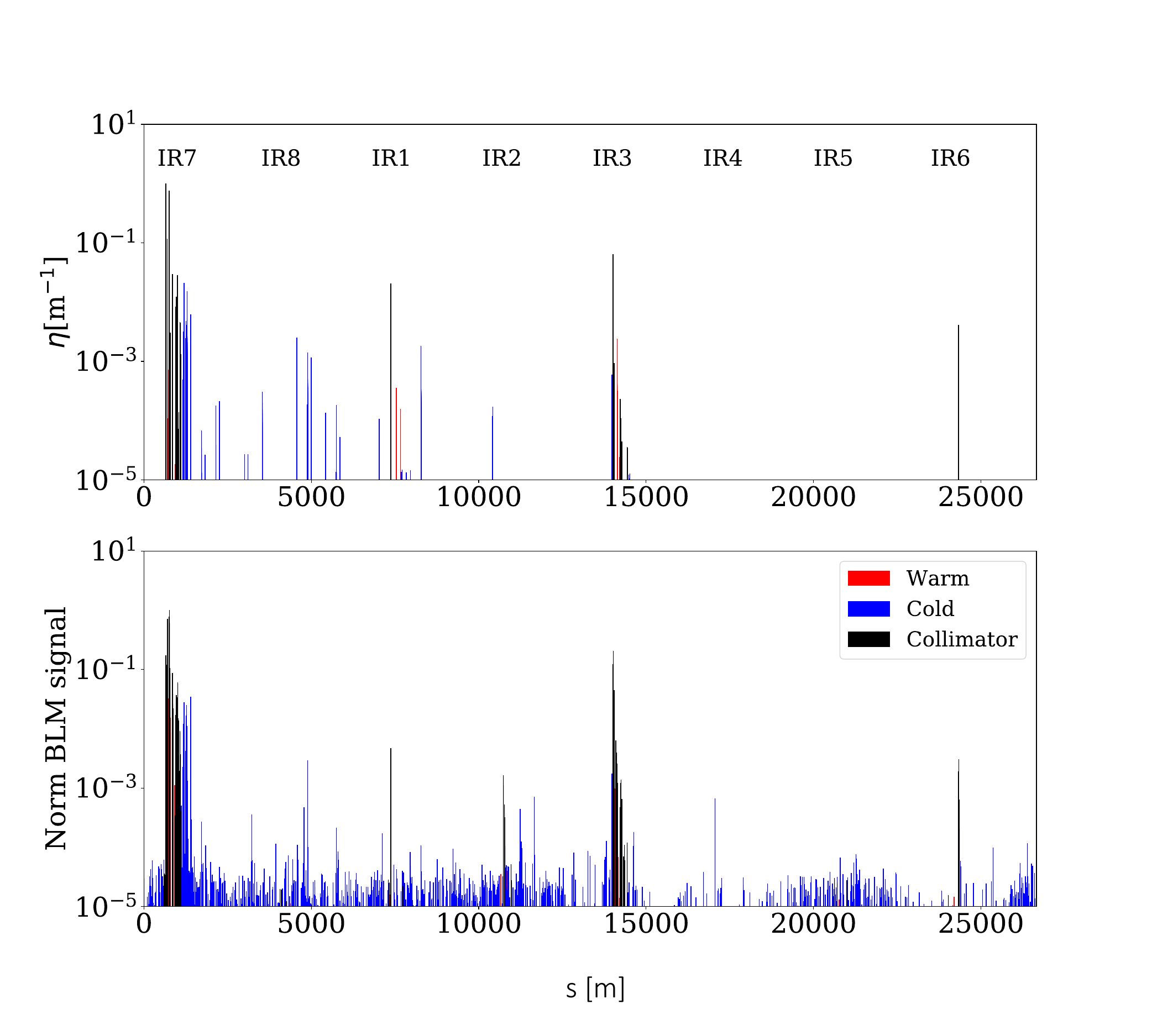}
    \caption{Simulated (top) and measured (bottom) FT B1 horizontal \Pb~ion beam loss map.}
    \label{Pb_FT} 
\end{figure}

Simulations have also been performed with the flat-top machine configuration with the optics, beam, and collimator settings summarised in Tables~\ref{BeamSettings} and~\ref{2018settings}. The comparison of simulated (top) and measured (bottom) loss maps for B1 in the horizontal plane is shown in Fig.~\ref{Pb_FT}.

A quite good qualitative agreement can be observed in the losses at the collimators except for the losses on the TCT in IR2 that are present in measurements, but not in simulations. This could potentially be explained by various imperfections, but a more detailed study would be required to verify this hypothesis. Cold spikes between IR7 and IR1 are observed in both simulations and measurements and the highest losses in the DS of IR7, are again well reproduced.

In comparison to the simulations performed with the small-\bs~optics, less losses are observed at the TCTs in IR1, IR2, and IR5 while in IR7 no significant changes are observed, as expected, since in IR7 the optics and collimator settings are the same for both optics.

ASBD failure measurements performed during the 2018 \Pb~ion commissioning have also been used to evaluate the \sixtrack-FLUKA coupling framework. In measurements only an event similar to the ASBD failures can be caused intentionally. However, in simulations also the SMPF has to be considered to evaluate the most pessimistic failure scenario in terms of beam losses around the machine and evaluate the amount of losses on the most sensitive components. The \sixtrack-FLUKA coupling was used to perform simulations of such failure scenarios, for the first time with ion beams, with subsequent analysis following the methodology described in Section~\ref{sec:level3}.

\begin{figure}[!htb]
    \centering
    \includegraphics*[width=9cm]{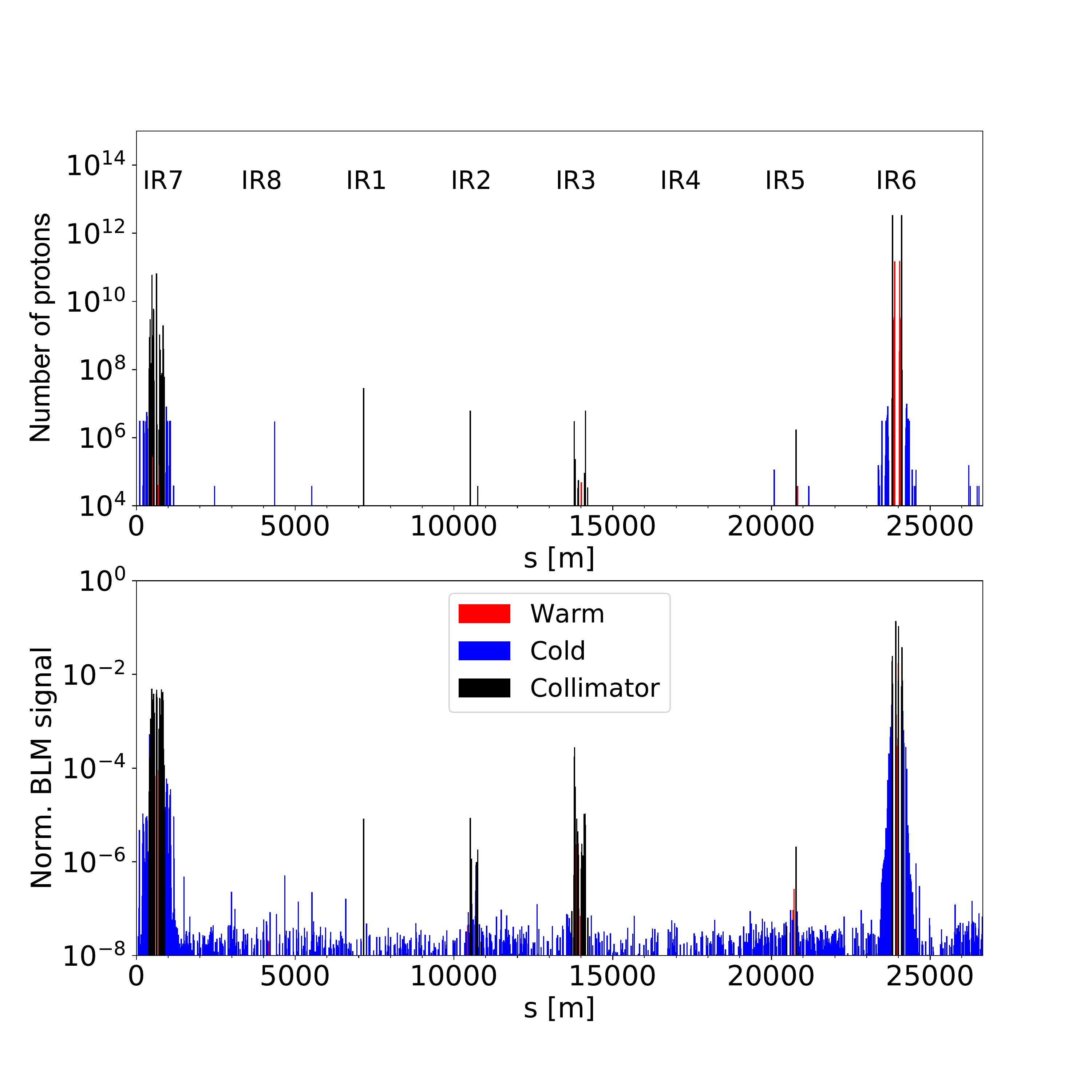}
    \caption{Simulated (top) and measured (bottom) loss map for an ASBD test with colliding beams during the 2018 \Pb~ion run commissioning.}
    \label{ASDLM} 
\end{figure}

As an example, in Fig.~\ref{ASDLM} the simulated (top) and measured (bottom) loss maps, resulting from a 2018 ASBD test performed with \Pb~ions and collision optics, is shown. It should be noted that in both plots, data from the two beams are superimposed, since the experimental procedure is performed for the two beams simultaneously. In addition, the simulated loss map is given in number of protons for comparison with collimator damage limit calculations performed for protons in~\cite{ElenaQ}. Good qualitative agreement is observed, with the main losses occurring this time in IR6, downstream of the extraction, with about one order of magnitude lower losses in IR7. Losses are observed also at the TCTs for B1 in IR1, for both beams in IR2, and at a lower level for B2 in IR5. Cold spikes are also observed in the arcs between IR8 and IR1 in both simulations and measurements. As discussed at the beginning of this section, the observed shifts in the longitudinal axis could be due to orbit and machine imperfections.

Serious damage can be caused to the TCTs if the primary and focused beam is intercepted (see~\cite{ElenaQ} for the case of protons). In order to ensure the protection of these collimators for the operation at small $\beta^*$ when tight TCTs settings are required, the fractional phase advance between the MKDs and these collimators have to be within~30$^\circ$ from 0$^\circ$ or 180$^\circ$ with proton beams~\cite{bruce17_NIM_beta40cm}. In Table~\ref{phaseadavnce} a summary of the phase advance between MKDs and TCTs for both beams in the 2018 \Pb~optics is presented. As can be seen, the TCTs in IP2 do not satisfy the requirements. Because of that, it was important to quantify the expected impacts for \Pb~ion beams, in order to conclude if whether the phase advance could be accepted. However, the \Pb~ion beams used at the LHC have a smaller bunch intensity and larger bunch spacing than the proton beams and the phase advance tolerances in such failure scenarios can be larger than for protons.

\begin{table}[!h]
\caption{\label{phaseadavnce} 2018 \Pb~ion run optics $\Delta \mu^{\rm TCT-MKD}$ summary for B1 and B2.}
\begin{ruledtabular}
\begin{tabular}{lccc}

 \textbf{IR}  &\textbf{$\Delta \mu ^{\rm TCTPH-MKD}_{\rm B1}$ }& \textbf{$\Delta \mu^{\rm TCTPH-MKD}_{\rm B2}$ }\\

\hline
1 &  176$^\circ$  & 151$^\circ$     \\
2 & 223$^\circ$    &  212$^\circ$    \\
 5 &    162$^\circ$  &    176$^\circ$  \\
\end{tabular}
\end{ruledtabular}
\end{table}
A sensitivity study was done, simulating the accidental losses as a function of the TCT half gap. This allows establishing the proper margins when defining the collimator settings, accounting for possible orbit errors during operation as in~\cite{bruce15_PRSTAB_betaStar}. 

\begin{figure}[!htb]
    \centering
    \includegraphics*[width=\columnwidth]{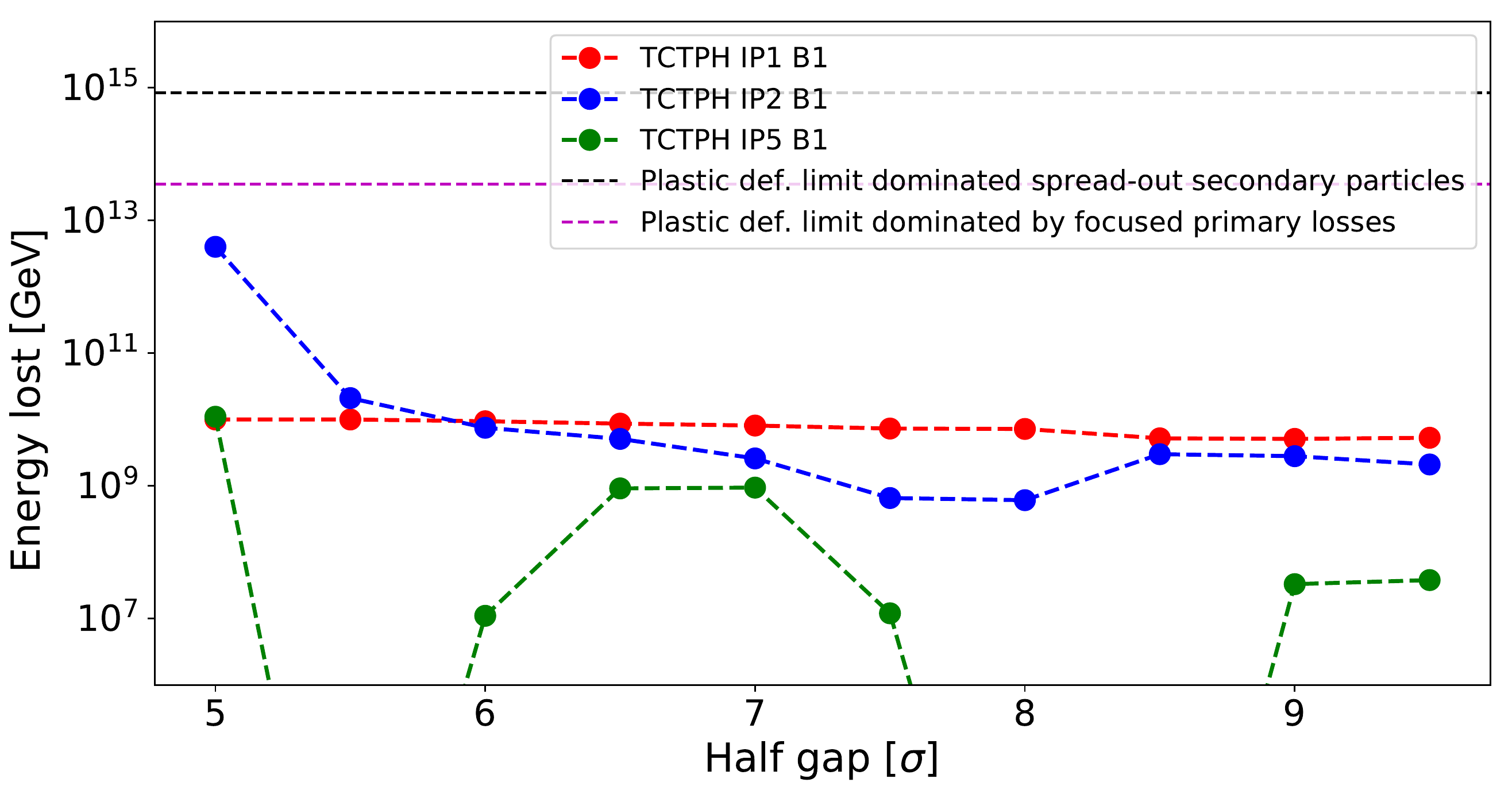}
    \includegraphics*[width=\columnwidth]{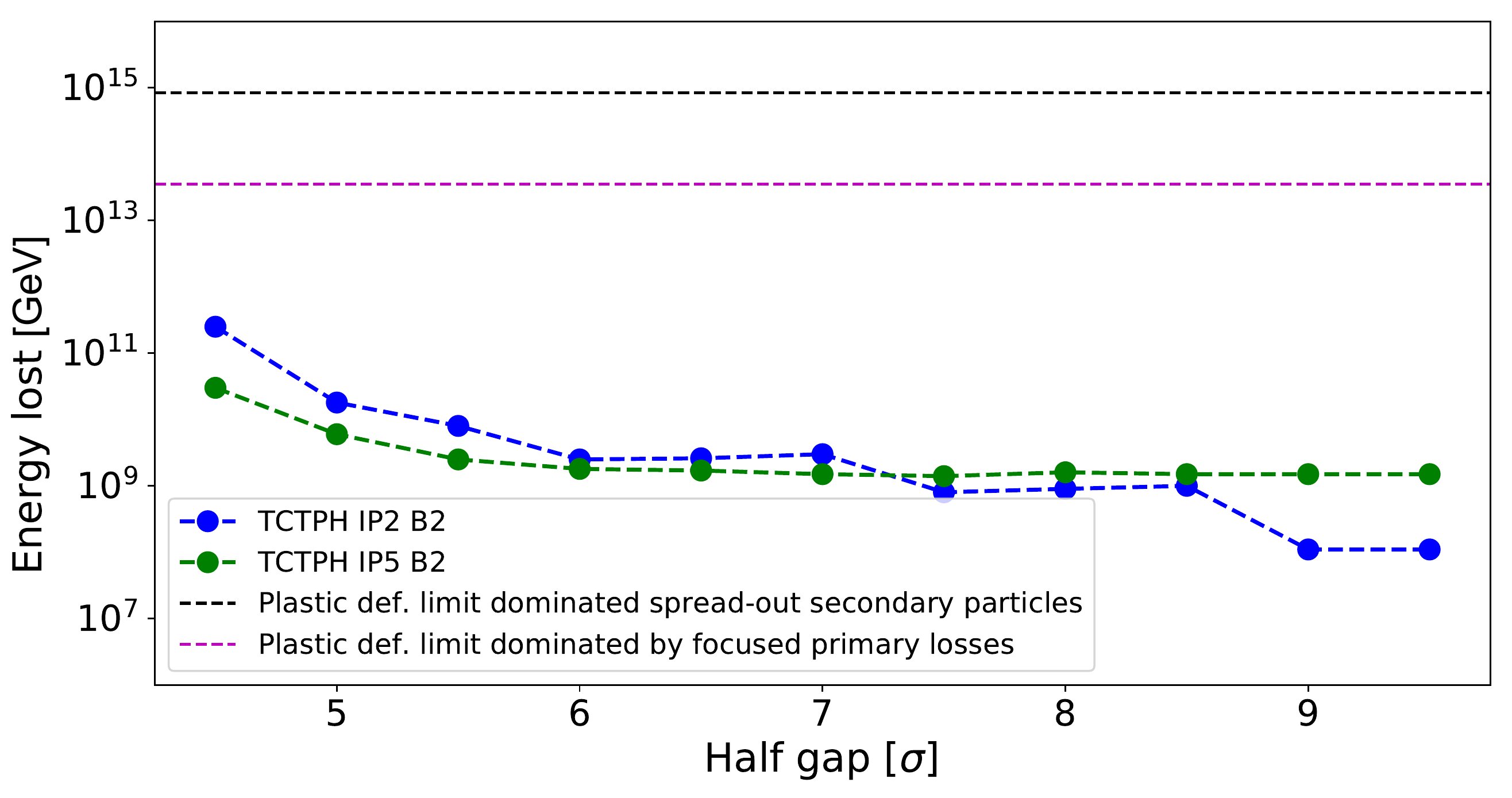}
    \caption{Losses at the TCTs at the higher luminosity experiments from simulation results of a SMPF mode scenario as a function of the TCTs half aperture for B1 (top) and B2 (bottom). Note that the discontinuity observed on the top plot green line is due to statistic limitations and indicates nearly no losses at this collimator. The energy lost by one primary \Pb~ion of 6.37 Z TeV in the simulations corresponds to 9.3$\times$10$^8$ GeV in this plot. For B2 no losses are observed in the TCT in IR1. }
    \label{ASD_scan} 
\end{figure}

In Fig.~\ref{ASD_scan} the results of these simulations are presented in the form of scaled energy lost at the TCTs in the different IRs as a function of the TCT half gap for B1 (top) and B2 (bottom), with all other collimators set according to the 2018 physics configuration in Table~\ref{2018settings}. The losses have been calculated using the set-up described in Section~\ref{sec:level3}.

In Fig.~\ref{ASD_scan}, the losses at the tungsten TCTs are compared with the estimated damage limits for protons~\cite{ElenaQ} obtained from detailed energy deposition and thermo-mechanical studies. The proton damage limit can be used in fairly good approximation, since the hadronic shower is very similar to the \Pb~ions, and the localised ionisation energy loss, which scales with $Z^2$ and is thus much higher for nuclear beams, would cause the peak energy deposition only for very small spot sizes. The purple dashed line indicates the plastic deformation limit for primary and focused beam losses, while the dashed black line corresponds to the plastic deformation limit for secondary spread-out beam. 

A flat dependence of the losses on the TCT setting can be observed for all the TCTs down to a minimum half gap of 5.5$\sigma$. Only for the horizontal TCT in IP2 we observe a steep rise when closing from 5.5$\sigma$ to 5$\sigma$. This means that up to this point all intercepted particles are from secondary spread-out beam, out-scattered at large amplitudes by the dump protection, as in Ref.~\cite{bruce17_NIM_beta40cm}, and the values are well below the damage limit from secondary beam halo. Notice that the discontinuity observed in Fig.~\ref{ASD_scan} for the TCTPH in IR5 (in green) could be explained by statistical limitations since a different initial beam distribution is generated for each bunch with only 12$\times$10$^4$ \Pb~ions. This indicates nearly no losses as the energy lost in the simulations by one primary \Pb~ion of 6.37 Z TeV corresponds to 9.3$\times$10$^8$ GeV in this plot. As can be seen in Fig.~\ref{ASD_scan} (bottom) for B2 losses are observed on the TCT in IR1.

From these studies we could conclude that the losses at the TCTs were well below the plastic deformation limits (indicated in black and purple in Fig.~\ref{ASD_scan}) with an operational margin of 3.5$\sigma$ between the operational TCT setting and the settings at which there is a risk for damage for the 2018 \Pb~ion collision optics, corresponding to the abrupt change on the slop of the curves of Fig.~\ref{ASD_scan}. The results show that the optics requirements, e.g. in terms of MKD-TCT phase advance, for \Pb~ion beams are less stringent than the ones for protons in Ref.~\cite{bruce17_NIM_beta40cm}. This is mainly due to the much lower bunch charge and larger bunch spacing, which both contribute to the lowering of the total impacting energy. 

\subsection{2015 \Pb~quench test}
\label{sec:2015_quench_test}
In a ``collimation quench test'', losses on the primary collimator are intentionally increased in a controlled way, causing losses in the cold DS magnets immediately downstream of the collimation insertion that could potentially cause a quench. The level of losses at which the quench occurs, or the largest losses without quench, can be compared with the estimated quench limit~\cite{Lechner2019}. These measurements provide almost ideal conditions for bench-marking simulations due to the resulting high signal-to-noise ratio and the control of the loss source. In addition, further information from the quench experiment performed in 2015 such as the power deposition in the SC magnet coils could be gathered by means of combining the measured loss rates with simulations which is of great importance to evaluate possible limitations in future runs with envisaged higher intensities. 
\begin{figure*}[!htb]
    \centering
     \subfloat[]{\includegraphics*[width=9.4cm]{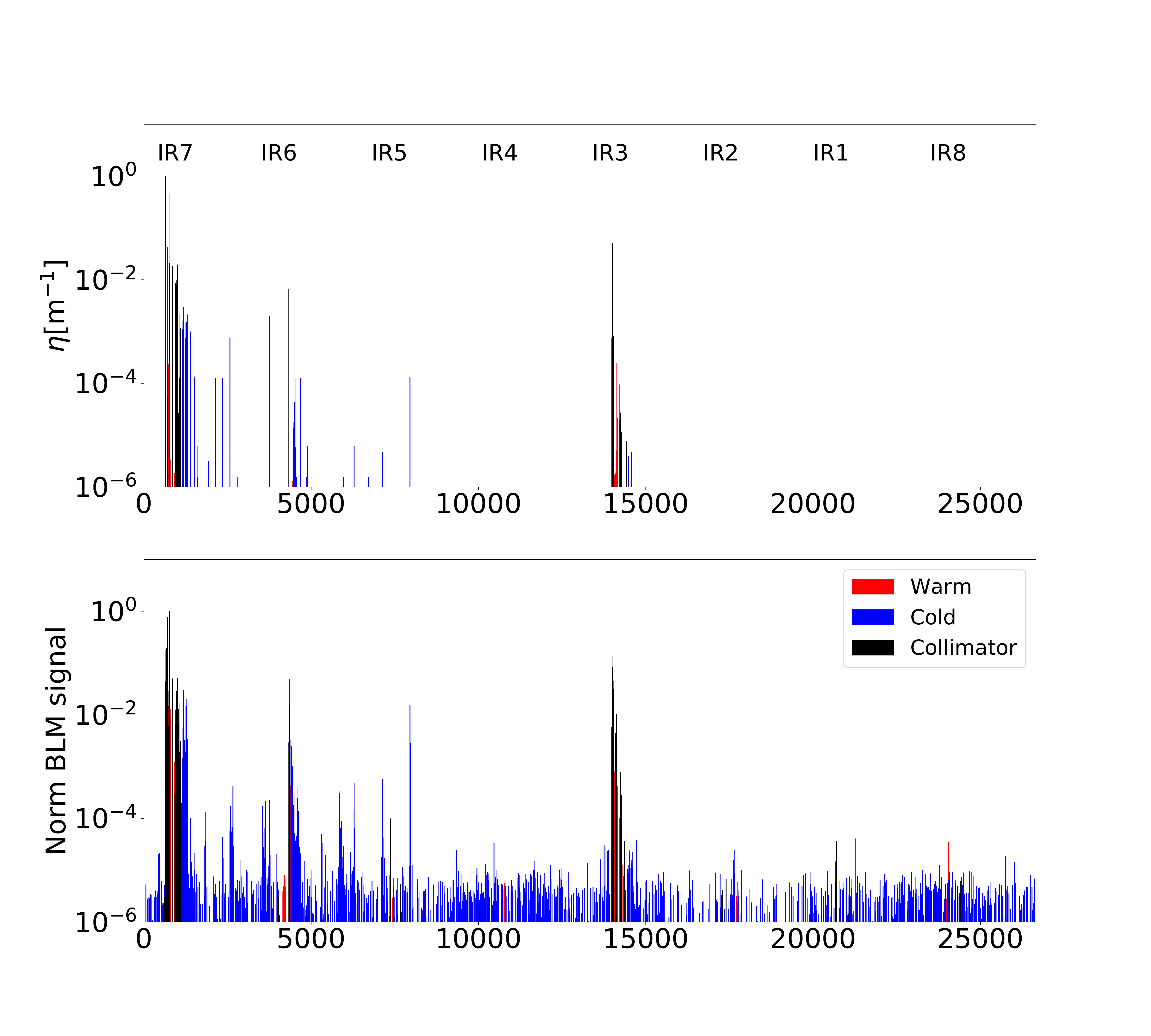} \label{2015_all}}
     \subfloat[]{\includegraphics*[width=9.2cm]{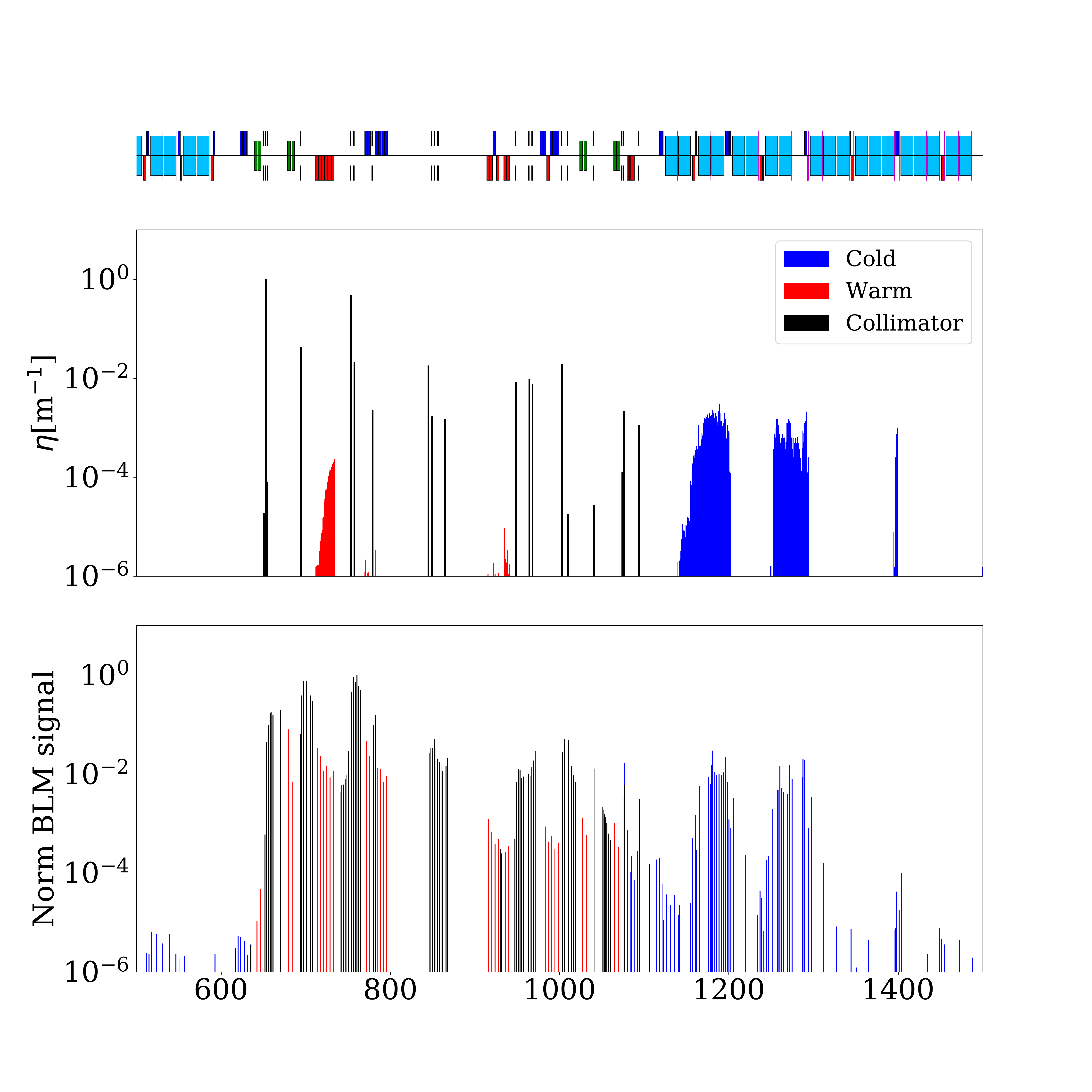}\label{2015_zoom}}
    \caption{Simulated (top) and measured (bottom) horizontal B2 full ring (a) and IR7 zoom (b) loss map for the 2015 \Pb~ion beam collimation quench test.}
    \label{LHC_quench_test} 
\end{figure*}
\begin{figure*}[!htbp]
    \centering
    \includegraphics*[width=0.9\textwidth]{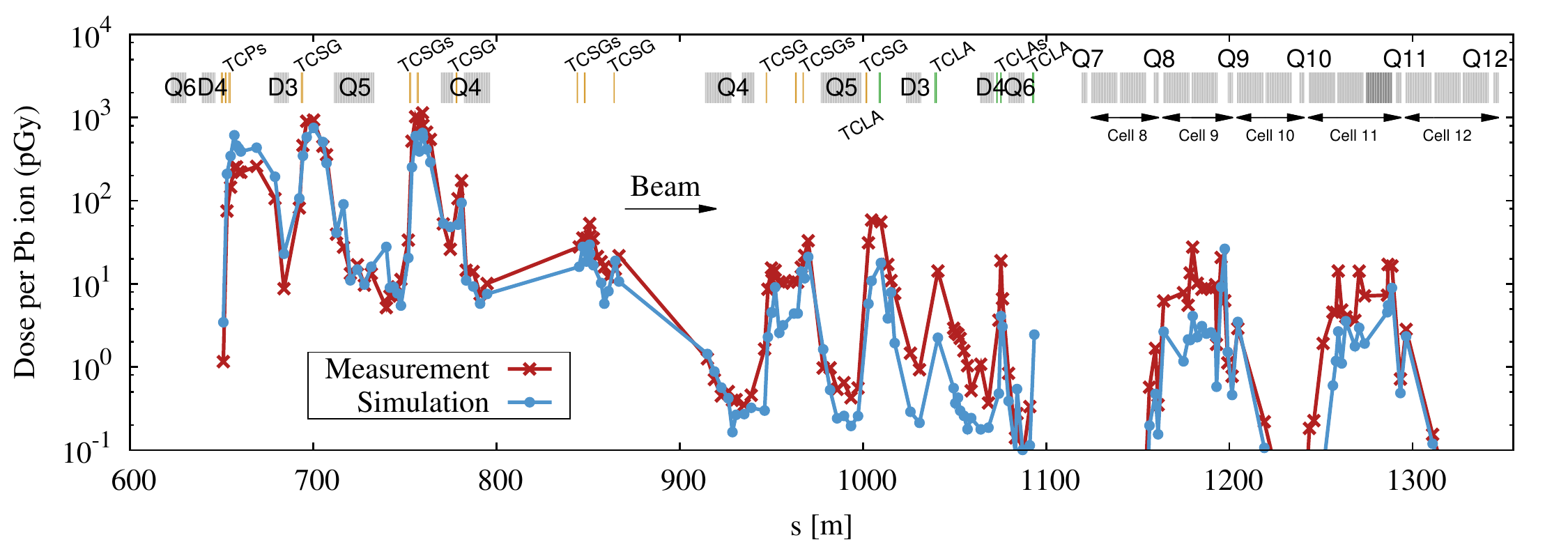}
    \caption{Comparison of simulated and measured BLM signals in the IR7 cleaning insertion, the adjacent DS (cells 8-11) and the first arc cell (cell 12). The beam direction is from the left to the right. The signals are expressed per \Pb~ion lost in the collimation system. The positions of primary and secondary collimators (TCPs and TCSGs), tungsten absorbers (TCLAs), dipoles (D3 and D4), and quadrupoles (Q4-Q12) are illustrated on the top. The statistical error of simulation results is generally smaller than 10\% for dose values above ∼10\,pGy but can be as large as a factor of a few for dose values below 10\,pGy.}
  \label{fig:qt_benchmark} 
\end{figure*}
Figure~\ref{LHC_quench_test} shows the \sixtrack-FLUKA coupling simulated (top) and measured (bottom) horizontal B2 \Pb~ion full ring (a) and IR7 zoom (b) loss map for the 2015 quench test configuration, shown in Table~\ref{BeamSettings}. Good qualitative agreement can be observed in the losses at the collimators except for the small amount of losses (about 10$^{-4}$) observed in the measured loss map on the TCT in IR5 that are not present in simulations. Cold spikes between IR7 and IR5 are observed in both simulations and measurements and the highest losses in the DS of IR7, are again well reproduced.

As anticipated in the introduction of this section, an improved quantitative comparison can be achieved by a two-step simulation combining tracking and detailed energy deposition studies. A similar approach was already adopted in Ref.~\cite{Lechner2019} for benchmarking BLM response simulations for proton collimation losses. In a first step, the \sixtrack-FLUKA coupling was used to generate the impact distribution of \Pb~ions at the primary collimator in IR7. In the second step, this impact distribution at the TCP was employed as initial distribution for the FLUKA energy deposition simulation with a detailed model of the IR7 geometry. In these simulations, the interaction of the particles with the primary collimator material and the subsequent propagation of residual fragments and electromagnetic and hadronic showers in the surrounding machine hardware is modeled. Secondary particles were transported until their kinetic energy fell below 1\,MeV (electrons, positrons), 100\,keV (photons, hadrons, muons) or 10$^{-5}$\,eV (neutrons). Detailed models of the loss monitors used in the LHC (nitrogen-filled ionisation chambers with a sensitive volume of about 1500\,cm$^3$) were included in the simulation setup to perform a quantitative comparison with the measured BLM signals. 

Since only a small fraction of secondary particles leak to superconducting magnets in the DS and arc, separate FLUKA shower simulations were carried out for the room temperature and the cold accelerator region. To enhance the statistical convergence in the latter case, the FLUKA simulation was split into two steps. In the first step, the transport of secondary particles and ion fragments emerging from interactions in collimators was suppressed if they could not reach the DS because of their magnetic rigidity. In this way, a representative distribution of secondary particles leaking to the cold region could be obtained since the computational time for simulating particle transport in the insertion region was significantly reduced. In the second step, the showers induced by the particles lost in cold magnets were simulated using the low transport thresholds mentioned above. More details about this simulation approach can be found in~\cite{Skordis:thesis}. The two steps were not necessary for obtaining BLM signals in the insertion region itself since a good statistical convergence could be achieved in a single step.

Figure~\ref{fig:qt_benchmark} compares simulated and experimental BLM signals in IR7, the neighbouring DS, and the first arc cell. The position of the primary and secondary collimators, tungsten absorbers and magnets is indicated on the top of the graph. The BLM signals are expressed per \Pb~ ion intercepted by the IR7 collimators and eventually lost in the machine. The simulation results represent the dose scored in the active gas volume of the BLM models. The experimental BLM dose values were derived by time-integrating BLM signals over the entire loss duration and by dividing the results by the number of ions lost from the beam. The latter was determined from beam current measurements, which provided a good estimate of betatron losses in IR7 since the  contributions of other loss mechanisms were much smaller. In total, about 1.4$\times$10$^9$ \Pb~ions were lost during the experiment. The measured BLM dose values were also corrected for the noise pedestal, which was obtained from a reference period without beam just after the experiment. The noise contribution was less than 1\% for the highest signals at primary and secondary collimators, and between 1\% and 10\% for the highest signals in the DS and arc.

In general, the simulated BLM pattern reproduces well the measured pattern, which spans several orders of magnitude in losses, over more than 600 m. Qualitative features like elevated BLM signals downstream of collimators and the two loss clusters in the DS are well reproduced. A good quantitative agreement, better than a factor of 2--3, can be observed for most BLMs at primary and secondary collimators. The simulation systematically underestimates, however, measured signals in the second half of the insertion region and in the cold accelerator region up to the Q11. The simulated signals in the DS are on average about four to five times lower than the measured ones. This can possibly be attributed to the assumed \Pb~impact distribution on TCPs and to machine imperfections that are not taken into account in the model. As shown in~\cite{Hermes:thesis}, the chosen maximum impact parameter of \Pb~ions on the TCPs influences the number of fragments leaking to the DS. In the present study, the maximum impact parameter was assumed to be 2\,$\mu$m. If the actual value was smaller, then this could have lead to a higher leakage as can be seen in Fig.~\ref{bscan}. In addition, machine imperfections like collimator tilts can also affect the collimation inefficiency. Imperfections could also be the main reason why the leakage of single diffractive protons to the DS was underestimated by about a factor of three in previous proton benchmark studies~\cite{Skordis2017,Lechner2019}. As shown in~\cite{Skordis:thesis}, the agreement for protons notably improved if primary collimators were assumed to be tilted. Qualitatively, a similar effect is expected for the leakage of \Pb~fragments. Despite the observed underestimation in Fig.~\ref{fig:qt_benchmark} the results are nevertheless considered remarkable, given the complexity of the simulation chain and the large variation of BLM signals, which span over many orders of magnitude.

\section{HL-LHC expected performance}
\label{sec:hl}
The LHC Run~1 (2010-2013)~\cite{BSalvachuaFerrando-IPAC13} and Run~2 (2015-2018)~\cite{NFusterMartinez-EVIAN2019} collimation operation was very satisfactory, with no collimation-induced quenches, however relevant upgrades are necessary to cope with the beams foreseen by the HL-LHC project~\cite{HLLHCtechDesRepo,Rossi:IPAC19-MOYPLM3}, which aims at increasing the integrated proton and heavy-ion luminosity collected by the LHC experiments by a factor 10 over twelve years. To reach this goal, it is, among others, foreseen to
use a higher proton-bunch population at injection of about ~2.3$\times$10$^{11}$ and to increase the number of bunches in the heavy-ion operation from 733, as used in 2018, to 1232, thanks to a shorter 50~ns bunch spacing~\cite{LIU-TDR}. 

\begin{figure}[!htb]
\begin{minipage}[!htb]{\columnwidth}
  \centering
  \includegraphics*[width=\columnwidth, trim=55mm 350mm 17mm  0mm, clip=true]{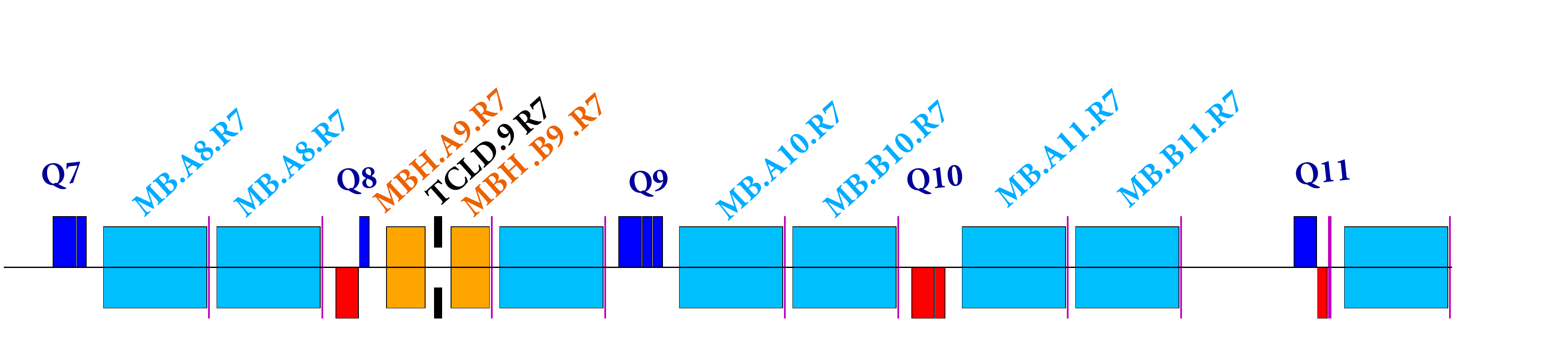}
  \includegraphics*[width=\columnwidth, trim=24mm 200mm 0mm  0mm, clip=true]{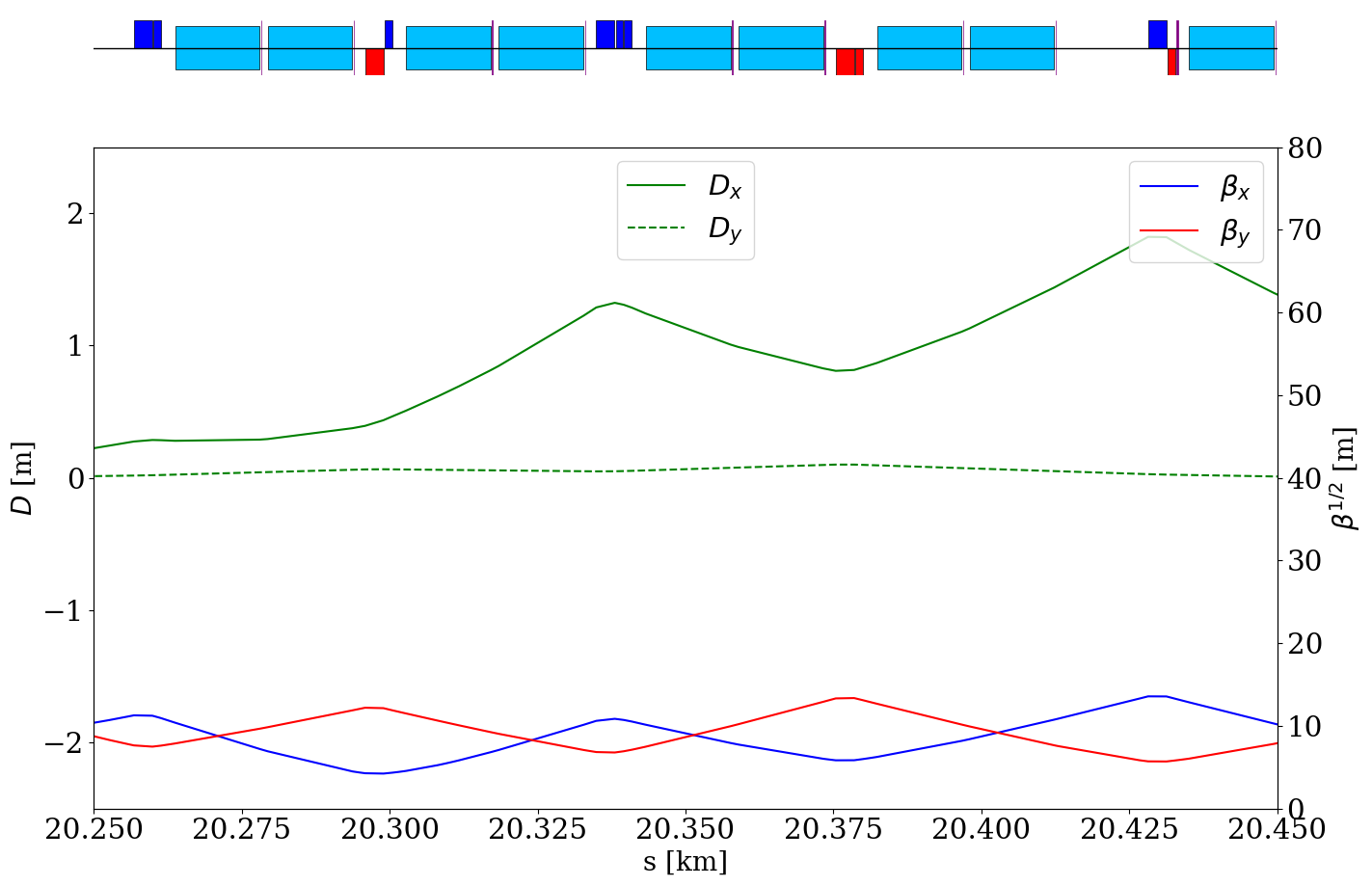}
\end{minipage}
\hfill
   \caption{Run~3 (top) and Run~2 (bottom) layouts of the IR7 DS for B1, with and without the new TCLD collimators. }
   \label{fig:IRxDSlayout}
\end{figure}

\begin{figure*}[!htbp]
    \centering
    \subfloat[]{\includegraphics*[height=6.8cm,width=8.6cm]{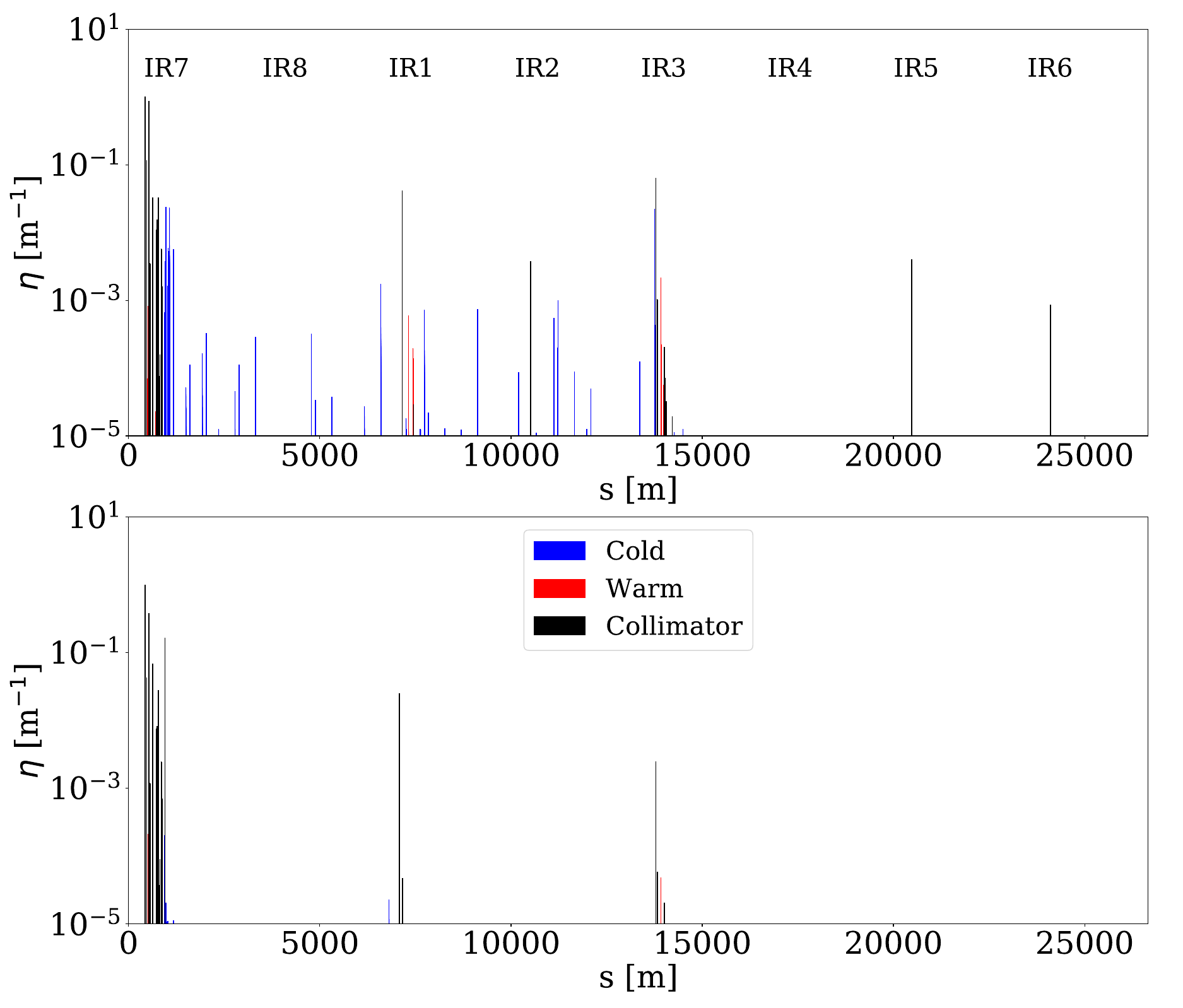} \label{HLLHCB1}}
     \subfloat[]{\includegraphics*[width=8.6cm]{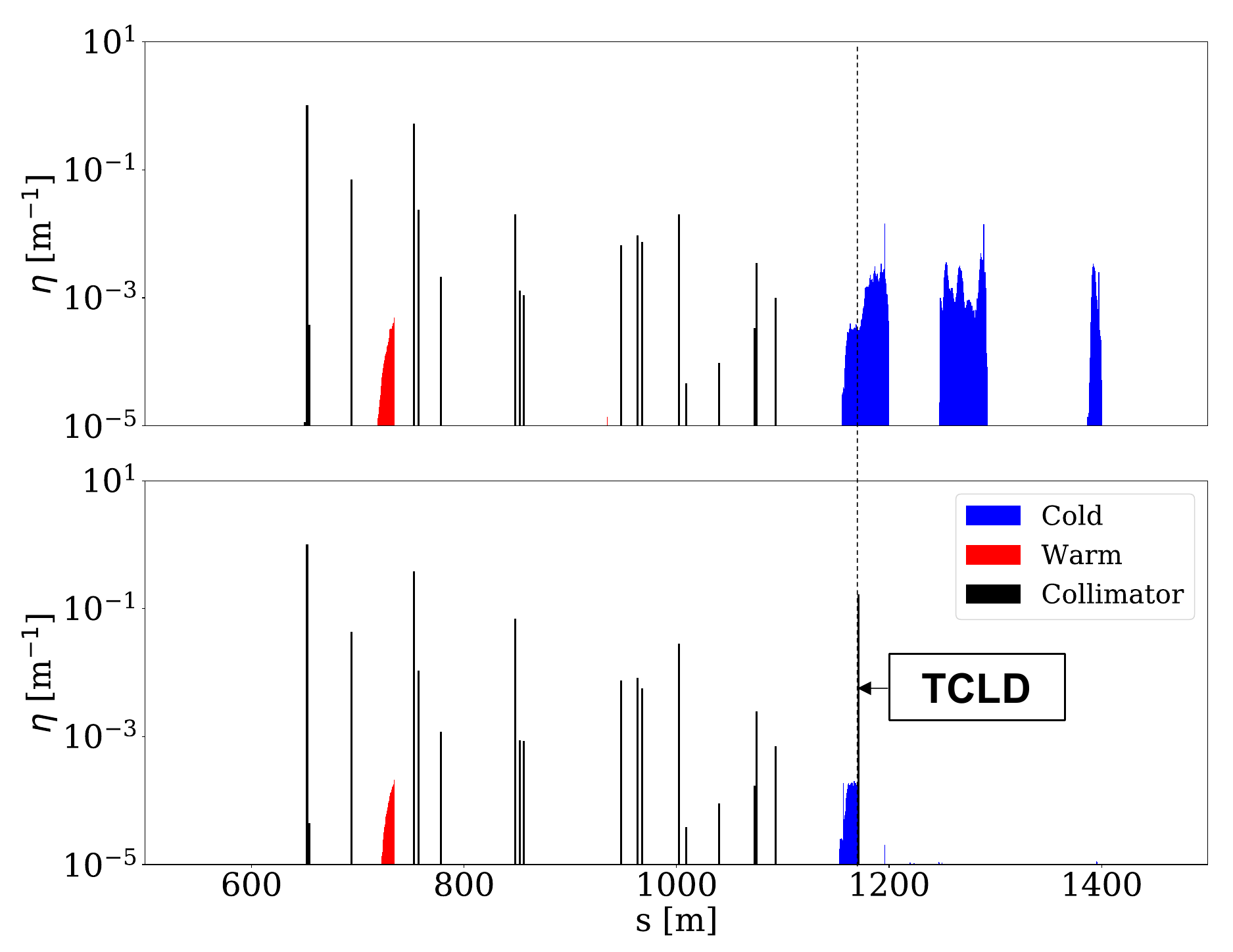}\label{HLLHCB1_zoom}}
    \caption{Horizontal B1 full ring (a) and IR7 zoom (b) loss map for the 2018 LHC (top) and HL-LHC (bottom) \Pb~ion operation layout.}
    \label{LHC_HLLHC} 
\end{figure*}

To cope with these high-intensity beams, some upgrades are planned for the collimation system with the aim of reducing the impedance contribution from the collimation system, to improve the cleaning efficiency in IR7, and to protect the experimental IRs from physics debris. The planned upgrade to improve the cleaning in IR7 is motivated mainly by predicted limitations in the HL-LHC collimation performance with heavy-ion beams. This limit is more constraining for \Pb~ion beams than for proton beams in spite of the factor~35 lower stored beam energy carried by the the \Pb~ion beams, due to the observed worsening of the IR7 cleaning performance (see Fig.~\ref{LM_example}). Based on energy-deposition studies~\cite{cristinaFLUKA}, the observed heavy-ion losses in simulations in the DS downstream of IR7 (see Fig.~\ref{LHC_HLLHC} top graph) scaled up to HL-LHC intensities are above the quench limit for lifetime drops below the design value of 12~minutes. 
To overcome this limitation, one new collimator per beam, called TCLD, will be installed in cell 9 on each side of IR7, as illustrated in the layouts in Fig.~\ref{fig:IRxDSlayout}. To make room for the TCLDs, a standard main dipole will be replaced by an assembly consisting of two shorter and stronger 11~T dipoles~\cite{savary15} (in orange in Fig.~\ref{fig:IRxDSlayout}) with the TCLD in the space in the middle. 

The performance of the collimation system for the new layout has been simulated using the \sixtrack-FLUKA coupling for \Pb~ion beams and the HL-LHC v1.2 optics. A 7~$Z$~TeV \Pb~ion beam with a normalised emittance of $1.49 \times 10^{-6}$~m~rad has been generated following the method described in Section~\ref{sec:level3}. The collimator settings considered in these simulations are summarised in Table.~\ref{HLLHC_settings}. Note that other collimator settings are being considered, but it is not expected that they will affect the conclusions of the studies presented in this paper. 
\begin{table}[!h]
\caption{\label{HLLHC_settings}HL-LHC \Pb~ion physics collimator settings for a normalised proton emittance of $\epsilon^P_N$ = 3.5~$\mu$m.}
\begin{ruledtabular}
\begin{tabular}{lcc}
\textbf{Collimator}& \textbf{IR} &\textbf{Half-aperture [$\sigma$]}\\\hline
TCP/TCSG/TCLA& 7& 6/7/10\\
TCP/TCSG/TCLA &3 &15/18/20  \\ 
TCTs &1/2/5/8  &10/10/10/15 \\ 
TCDQ / TCSP&6 &9/9 \\ 
TCL 4/5/6&1/5  &12 \\
TCLD &7 &14\\
\end{tabular}
\end{ruledtabular}
\end{table}

\begin{figure*}[!htb]
    \centering
     \subfloat[]{\includegraphics*[height=6.8cm,width=8.6cm]{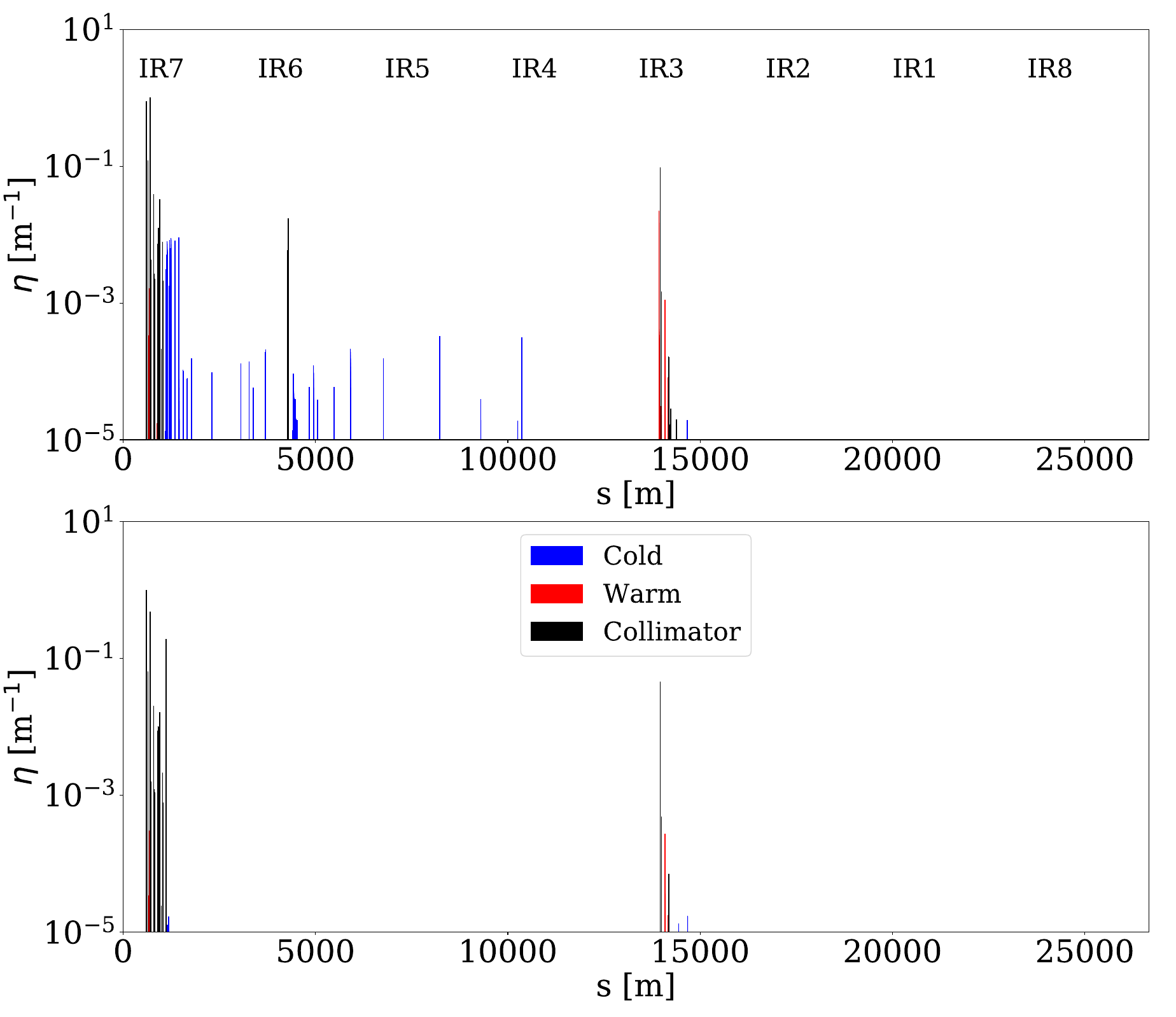} \label{HLLHCB2}}
     \subfloat[]{\includegraphics*[width=8.6cm]{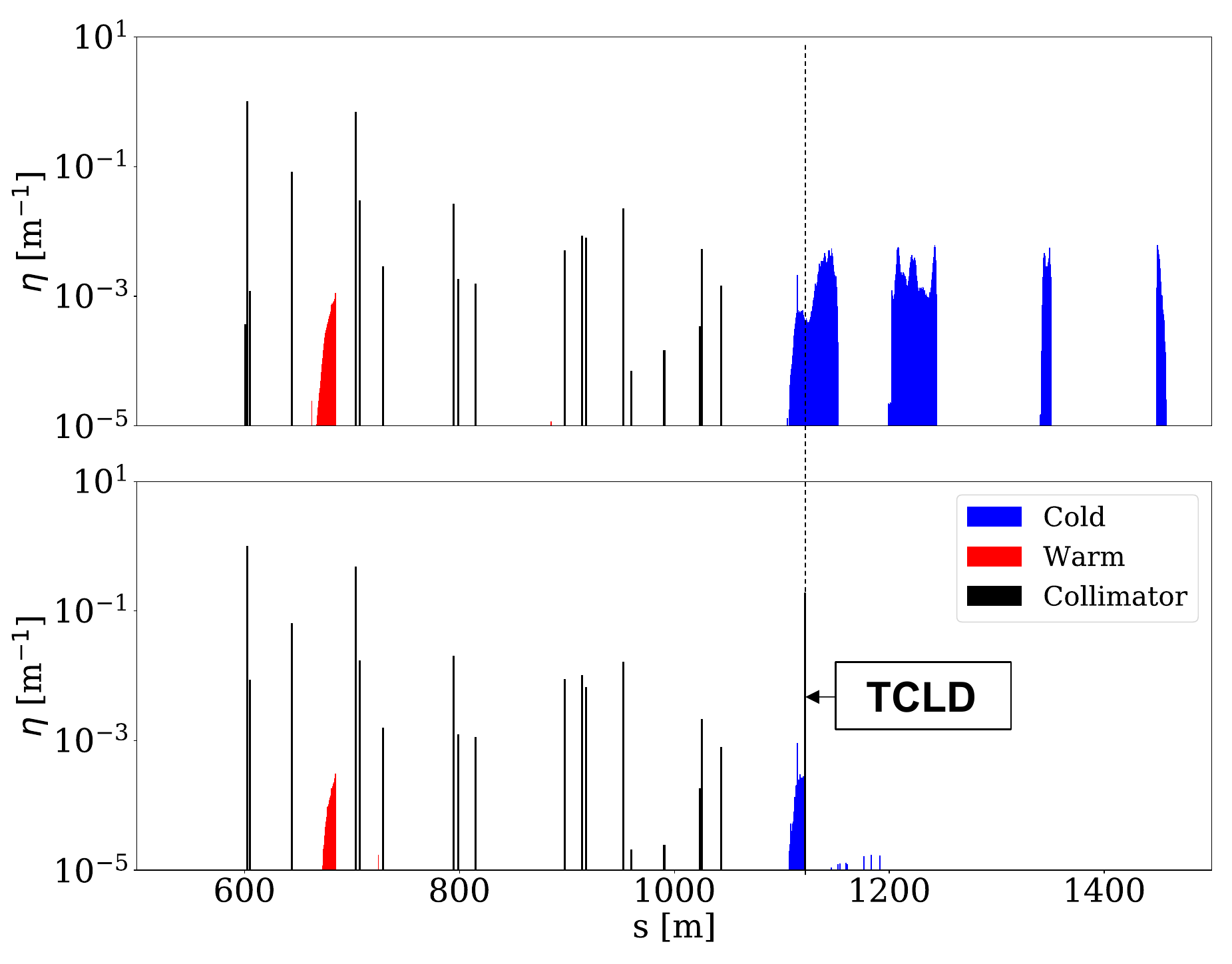}\label{HLLHCB2_zoom}}
    \caption{Horizontal B2 full ring (a) and IR7 zoom (b) loss map for the LHC (top) and HL-LHC (bottom) \Pb~ion operation layout.}
    \label{LHC_HLLHC_B2} 
\end{figure*}

In Fig.~\ref{LHC_HLLHC} and Fig.~\ref{LHC_HLLHC_B2}, the simulated horizontal LHC (top) and HL-LHC (bottom) loss maps are shown for B1 and B2, respectively. The full ring loss maps are presented in Fig.~\ref{HLLHCB1} and Fig.~\ref{HLLHCB2} and a zoom of IR7 is shown in Fig.~\ref{HLLHCB1_zoom} and Fig.~\ref{HLLHCB2_zoom} for B1 and B2, respectively.
With the TCLD in place in IR7, the cleaning all along the machine is greatly improved, as can be seen in both Figs.~\ref{LHC_HLLHC} and~\ref{LHC_HLLHC_B2}. For B1, collimator losses are localised in IR7, IR3 and at the horizontal TCTPH in IR1. High losses in the TCTPH in IR1 are envisaged at the same level as observed in the 2018 \Pb~ion run. The mitigation strategies described in Section~\ref{sec:benchmark} could in principle be used. For B2, losses at collimators are only observed in IR7 and IR3. It should be noted especially that cold spikes all along the ring are almost fully suppressed, in particular, in IR7 where the limiting cluster losses defined in Fig.~\ref{LM_example} as DS2 and DS3 are completely mitigated and in DS1 the losses are reduced by more than a factor 10. This is because all these suppressed losses were caused by ion fragments with a magnetic rigidity different from the main beam. These fragments are efficiently intercepted by the TCLD that is in a location with a significant dispersion. Detailed energy deposition studies have shown that the remaining cold losses occurring upstream of the TCLD are not limiting the achievable beam intensity~\cite{cristinaFLUKA}.
    
\section{Conclusions}
\label{sec:conclusions}
A good understanding of Heavy-ion collimation is essential to devise collimator settings providing safe operation, which is not interrupted by beam dumps or magnet quenches. In the short heavy-ion runs, the machine availability is crucial and every fill contributes significantly to the total integrated luminosity. Moreover, the collimation set-up and validation has to be done only once at the start of the run to avoid interruptions of the operation for physics. It is then very important to have accurate models to identify potential performance bottlenecks beforehand and to be able to react quickly in case of unexpected issues during the operation.

In the last ten years, a big effort has been made to improve the accuracy of the heavy-ion collimation simulation tools, combining reliable fragmentation models and precise tracking codes.
This is a complex task because it requires putting together different tools  optimised for different purposes.
This effort has resulted in the development of the \sixtrack-FLUKA coupling framework for heavy-ion beams, which has demonstrated its increasing reliability as a guide to understanding the origin and location of the losses in different scenarios such as standard collimation cleaning or dump failure scenarios. 

For the 2018 \Pb~ion run commissioning, the simulations were applied to mitigate losses through optimisation of the collimator settings, and the results were confirmed by experimental studies. In general, very good qualitative agreement is observed in the losses at the collimators between our simulations and the measurements. The response of the BLMs could explain some of the observed discrepancies as well as machine imperfections, which are not taken into account in the simulations. In addition, the bottleneck location of losses in cold magnets is very well reproduced in the DS of IR7. Using these developed tools, effective mitigation strategies can be formulated and tested before implementation, given the predictive power of numerical simulations as shown for the 2018 \Pb ion run. The qualitative agreement between measurements and simulations is improved when detailed FLUKA energy deposition studies are performed including the complete geometry of the accelerator and BLMs, and a quantitative agreement is found within a factor of a few to over several hundreds of meters throughout the whole IR7. Given the complexity of the simulations, the fact that the losses span many orders of magnitude, and the many unknown error sources, we consider this a very good agreement. These results demonstrate the maturity of the simulation chain and give confidence in the simulations of future machine configurations. 

The performance of the HL-LHC project collimation layout has been evaluated with the the coupling simulation tool. The results show a very significant improvement with respect to the current LHC layout and no expected limitations for the heavy-ion run are predicted after the deployment of new DS collimators, whose location was optimised thanks to the new tools described in this paper.

\begin{acknowledgments}
The authors would like to thank colleagues in the collimation team for valuable discussions as well as the LHC operation team for their help in the acquisition of the loss maps analysed in the paper. We express also our gratitude to J. Jowett and M. Schaumann for helpful discussions.
\end{acknowledgments}

\appendix
\section{Tracking maps}
In the following text, we describe a generalised Hamiltonian for multi-isotopic particle beams used to derive the thin-lens symplectic tracking maps implemented in \sixtrack.

Consider the trajectory of an arbitrary particle of \mbox{rest mass $m$} and charge $Ze$ (with the charge multiplicity $Z$ and elementary charge $e$) moving at the relative speed $\beta=\frac{v}{c}$ through a dipolar magnetic field $B$. The trajectory is bent by a bending radius $\rho$, which is related to the magnetic field, the particle momentum, and charge as
\begin{align}
  B  \rho = \frac{P}{Ze} \, .
\end{align}
The particle momentum can be written as $P = m \beta c \gamma $ with $\gamma$ the relativistic normalised energy. The bending radius $\rho_0$ of the reference particle, with its physical properties defined by the parameters $m_0,Z_0,\beta_0$, is related to $\rho$ as follows
\begin{align}
\frac{\rho}{\rho_0} = \frac{(1+\delta)}{\chi} \, , \quad
\chi = \frac{m_0}{m} \frac{Z}{Z_0} \, , \quad  (1+\delta) = \frac{\beta \, \gamma}{\beta_0 \, \gamma_0} \, .  \label{rhorho0}
\end{align}
The quantity $\chi$ defines the mass-to-charge ratio of the ion relative to the reference particle and the quantity $\delta$ is the relative offset of the normalised relativistic momentum.

Elementary transformations of Eqs.~\eqref{rhorho0} show that $\delta$ in the multi-isotopic case is not the well-known relative momentum offset, but the relative momentum offset per mass unit
\begin{align}
\delta =  \frac{P}{m}\frac{m_0}{P_0}-1 \, .
\label{delta}
\end{align}
For the case of heavy-ions, the relative momentum offset of Eq.~\eqref{delta} can be larger than two orders of magnitude. Both $\chi$ and $\delta$ quantify the dispersive offset of the particle trajectory acquired after interaction with the collimator material. Note that for the mono-isotopic case when the same ions of the main beam are produced, $m\rightarrow m_0$ and $Z\rightarrow Z_0$, the two Eqs.~\eqref{rhorho0} and \eqref{delta} yield to the well-known expressions in which $\delta$ is the relative offset of the full momentum. 
 
Consider a physical system described by the canonical co-ordinates $\mathbf{p}, \mathbf{q}$ with $\mathbf{p}=\{p_x,p_y,p_z\}$ and  $\mathbf{q}=\{x,y,z\}$. After the transformation of the independent variable from $t$ to the path length $s(t)$, the accelerator Hamiltonian for the set of canonical variables $(x,p_x), (y,p_y), (-t, E)$ is given by~\cite{courant_snyder}
\begin{widetext}
\begin{align}
&H = -p_z = -Ze A_z - \sqrt{ \frac{(E-Ze\phi)^2}{c^2} - m^2c^2 - (p_x - Ze  A_x)^2-  (p_y -Ze  A_y)^2 } \, ,\label{eq:h001}
\end{align}
\end{widetext}
where $\phi$ is the scalar potential, $A_i$ the electromagnetic vector potential, defining the magnetic field vector \mbox{$\mathbf{B} = \nabla \times \mathbf{A}$} and the canonical momenta $p_i$ are defined as
\begin{align}
  p_i = m \gamma \dot{q_i} + Ze  A_i \, . 
\end{align}
The total particle energy $E$ in the square root of Eq.~\eqref{eq:h001} is a very large quantity. In order to solve analytically complicated vector potentials, the Hamiltonian should be expanded and this requires the dynamic variables in the square root to be small. The following substitutions of $E$, $p_i$, $A_i$ and $H$ serve the purpose of obtaining small dynamic variables in the square root of Eq.~\eqref{eq:h001}, while maintaining the validity of Hamilton's equations:
\begin{alignat}{4}
p_i &\rightarrow \tilde{p}_i = \frac{p_i}{P_0} \, \frac{m_0}{m} \quad \quad &H &\rightarrow \tilde{H} = \frac{H}{P_0}  \frac{m_0}{m} \, , \\
Ze\,A_i &\rightarrow \chi a_i = \chi \frac{Z_0e A_i}{P_0}  \quad \quad &E &\rightarrow \tilde{E} = \frac{E}{P_0} \, \frac{m_0}{m} \, . 
\end{alignat}
The normalisation with respect to the mass is essential to fulfil the requirement of obtaining small quantities in the square root of Eq.~\eqref{eq:h001} because the masses of the different ions produced by the interaction with collimators can differ significantly from the mass of the main ion beam. Note that the definition of the normalised vector potential $a_i$ is identical to the definition for the mono-isotopic case~\cite{wolski}. Instead of incorporating it into the definition of $a_i$, the magnetic rigidity change for isotopes different from the reference particle is taken into account by the additional factor $\chi$. This allows the usage of well-known vector potentials from the derivation of the mono-isotopic tracking maps~\cite{dipole,heinemann}. 

Expressed in terms of the new co-ordinates, and assuming that a gauge can be found, such that $\phi=0$, and using the relativistic energy-momentum relation and Eq.~\eqref{delta}, the Hamiltonian can be written as

\begin{align}
\tilde{H} = - \chi a_z - \sqrt{(1+\delta)^2  - (\tilde{p}_x - \chi a_x)^2 - (\tilde{p}_y-\chi a_y)^2 }  \, . 
\label{h001_2}
\end{align}
This equation is similar to the standard expression used in~\cite{heinemann}. It should, however, be kept in mind that the quantities $p_i$, $\tilde{H}$ and $\delta$ are defined differently.

In order to also describe the longitudinal particle motion (e.g. the synchrotron motion) by small quantities, another transformation is required that can be obtained by means of a transformation of the canonical variables provided by a generating function of second type~\cite{wolski}
\begin{align}
F_2 = x \, P_x + y \, P_y + (s-\beta_0 \, c \, t) \, \left(p_z + \frac{E_0}{\beta_0 \, P_0 \, c} \right)\, . 
\end{align}
The old ($\tilde{p}_i$, $q_i$) and new ($P_i$, $Q_i$) co-ordinates, as well as the old ($\tilde{H}$) and new ($K$) Hamiltonian are related by the following relations
\begin{align}
\tilde{p}_i = \PD{F_2}{q_i} \quad \quad Q_i = \PD{F_2}{P_i}  \quad \quad K = \tilde{H} + \PD{F_2}{s} = \tilde{H}+p_z  \, .
\end{align}
The transformed variables $(X,P_x), (Y,P_y), (z, P_z)$ are then defined as follows
\begin{alignat}{5}
X  &= x  \, ,          \quad \quad  &Y   &&= y  \, ,           \quad \quad &z   &&= \phantom{m} s - \beta_0 ct \, ,  \\ \label{eq:sigmadefinition}
P_x&= \tilde{p}_x  \, , \quad \quad  &P_y &&= \tilde{p}_y \, ,  \quad \quad &p_z &&= \frac{\frac{m_0}{m} \, E - E_0}{\beta_0 P_0 c} \, .
\end{alignat}
Including a last transformation for convenience $P_i \rightarrow p_i$, $K \rightarrow H$, the final multi-isotopic Hamiltonian in a straight co-ordinate system yields
\begin{align}
  H = p_z - \sqrt{(1+\delta)^2 - (p_x - \chi a_x)^2 - (p_y-\chi a_y)^2} - \chi a_z \, . 
\end{align}
The new co-ordinate $z$ describes the difference in arrival time with respect to the reference particle. The quantity $p_z$ is the canonical conjugate of $z$.

To describe particle motion in an accelerator it is convenient to transform the straight co-ordinate system into a curved one, with the new set of variables $(X,P_x), (Y,P_y), (s, P_s)$. In a co-ordinate system horizontally bent by a constant radius \mbox{$\rho_0 = 1/h_x$}, the Hamiltonian becomes
\begin{widetext}
\begin{align} 
H=p_{z} -(1-h_x(s)x)\left ( \sqrt{(1+\delta)^2-(p_x-\chi a_x(s))^2-(p_y-\chi a_y(s))^2}+\chi a_s(s)\right )\, ,
\label{Hamiltonian_exact}
\end{align}
\end{widetext}
where $a_s$ is the vector potential in the curvilinear reference co-ordinates. The full derivation can be found in~\cite{Hermes:thesis}. In the mono-isotopic limit $m \rightarrow m_0$ and $Z \rightarrow Z_0$ the multi-isotopic Hamiltonian becomes the standard Hamiltonian presented in~\cite{proceedingsCAS,wolski}.

Depending on the complexity of the electromagnetic field of the beamline element and the corresponding boundary conditions it is useful to expand the square root of the Hamiltonian in  Eq.~(\eqref{Hamiltonian_exact}) in $\frac{(p_x-\chi a_x)^2 + (p_y - \chi a_y)^2}{(1+\delta)^2}$ to first order, as it is done for mono-isotopic beams in~\cite{courant_snyder}, and the Hamiltonian becomes
\begin{widetext}
\begin{align} 
H \approx p_{z}-(1-h_x(s)x)\left [ (1+\delta)\left (1-\frac{1}{2}\frac{(p_x-\chi a_x(s))^2+(p_y-\chi a_y(s))^2}{(1+\delta)^2}\right )+\chi a_s(s) \right ]\, .
\label{Hamiltonian_expanded}
\end{align}
\end{widetext}
The accuracy of the tracking maps derived using the expanded Hamiltonian in Eq.~\eqref{Hamiltonian_expanded} was studied for the drift space in the mono-isotopic case. The results using the exact Hamiltonian are in very good agreement if $p_x$ and $p_y$ are small and significant differences arise only if these values are so large that the particles would be lost in the magnet aperture after only a few meters~\cite{Hermes:thesis}. For the drift-space element, both tracking maps derived from the exact and the expanded multi-isotopic accelerator Hamiltonian in the thin-lens approximation were implemented in \sixtrack. The user can chose which option should be used depending on the requirements on the simulations precision and time. No significant increase in CPU time is expected if the number of simulated turns is smaller than 10$^5$~\cite{Fjellstrom:1642385}. Based on simulation studies a value of 700 turns have been defined and used in all cases presented in this paper, for which all simulated particles are lost in the ring. For the other beamline elements only the expanded Hamiltonian tracking maps were implemented because under the thin-lens approximation higher order terms cancel and the resulting tracking maps from one or the other Hamiltonian are equivalent. The symplecticity of the tracking maps was demonstrated by the means of the Jacobian matrix and the details of the derivation can be found in~\cite{Hermes:thesis}. 

In the following, the developed tracking maps from the approximated multi-isotopic Hamiltonian for the different accelerator elements are described. Notice that in \sixtrack instead of the transverse canonical momenta $p_x$ and $p_y$ the evolution of $x'$ and $y'$ is computed. As examples, the tracking maps implemented in \sixtrack for a thin-lens kicker dipole and the quadrupole are also presented. Further details on the implementation in \sixtrack are given in~\cite{Hermes:thesis}.

\subsection{Drift space}

A drift space is defined by the absence of electromagnetic fields $a_i=0$. The ideal trajectory is not bent, thus $h_x=0$ and the expanded Hamiltonian yields
\begin{align}
  H \approx p_\sigma - \delta +\frac{1}{2} \frac{p_x^2 +p_y^2}{(1+\delta)} .
\end{align}
The resulting tracking maps are independent of the ion species and thus identical to the mono-isotopic case. The resulting tracking maps are
\begin{equation}
\begin{pmatrix}
x^{F}\\y^{F}\\z^{F}
\end{pmatrix}
=
\begin{pmatrix}
x^{I}+\frac{p_x^{I}}{1+\delta}L\\y^{I}+\frac{p_y^{I}}{1+\delta}L\\z^{I}-L\frac{\beta_0}{\beta_z}\left ( 1+\frac{1}{2}\frac{(p_x^I)^2+(p_y^I)^2}{(1+\delta)^2}\right)
\end{pmatrix}
\end{equation}
\begin{equation}
\begin{pmatrix}
p_x^{F}\\p_y^{F}\\p_z^F
\end{pmatrix}
=
\begin{pmatrix}
p_x^{I}\\p_y^{I}\\p_{z}^{I}
\end{pmatrix}
\end{equation}

\subsection{Dipole Magnet}
\subsubsection{Bending Dipole}
Using the vector potential of a bending dipole derived in~\cite{dipole}
\begin{align}
a_x = a_y = 0\, , \quad \quad a_s = k_0 \, \left( x + \frac{h_x x^2}{2} \right) \, ,
\end{align}
the expanded Hamiltonian for a horizontal bending dipole magnet with the normalised strength $k_0= \frac{B_y \, Z_0 e}{P_0}$ and \mbox{$h_x \neq 0$} is given by
%
%
\begin{align}
  H \approx p_\sigma  &- (1+h_x  x) \, (1+\delta) + \notag \\  &+ \frac{1}{2} \, \frac{p_x^2 + p_y^2}{(1+\delta)} + \chi  k_0 \, \left(x + \frac{h_x  x^2}{2} \right) \, . \label{dipoleH}
\end{align}
The tracking map for a dipole of length $L$ in thin-lens approximation $k_0 L \rightarrow 0$ derived using Hamilton's equations are

\begin{equation}
\begin{pmatrix}
x^{F}\\y^{F}\\z^{F}
\end{pmatrix}
=
\begin{pmatrix}
x^{I}\\y^{I}\\z^{I}
\end{pmatrix}
\end{equation}

\begin{equation}
\begin{pmatrix}
p_x^{F}\\p_y^{F}\\p_z^F
\end{pmatrix}
=
\begin{pmatrix}
p_x^I+L[h_x(1+\delta)-k_0\chi(1+h_x x^I)]\\p_y^{I}\\p_{z}^{I}
\end{pmatrix}
\end{equation}
\subsubsection{Kicker Dipole}
The magnetic kicker dipole provides a transverse magnetic field, similar to the bending dipole, but the reference orbit is not bent ($h_x = 0$). Kicker dipoles are used to control the orbit in a machine. From the Hamiltonian in Eq. \eqref{dipoleH} with $h_x=0$, the resulting tracking maps are
\begin{equation}
\begin{pmatrix}
x^{F}\\y^{F}\\z^{F}
\end{pmatrix}
=
\begin{pmatrix}
x^{I}\\y^{I}\\z^{I}
\end{pmatrix}
\end{equation}

\begin{equation}
\begin{pmatrix}
p_x^{F}\\p_y^{F}\\p_z^F
\end{pmatrix}
=
\begin{pmatrix}
p_x - k_0 \, \chi \, L\\p_y^{I}\\p_{z}^{I}
\end{pmatrix}
\end{equation}
Taking into account that $x'=\frac{p_x}{(1+\delta)}$, the transformation of $x'$ yield 
\begin{equation}
    (x')^{F}=(x')^{I}-k_0L\frac{\chi}{(1+\delta)}
\end{equation}
\subsection{Quadrupole}
The quadrupole magnets are used to provide focusing in order to confine the transverse dimension of the beam. The vector potential of a horizontal or vertical quadrupole magnet in normalised co-ordinates is given by
\begin{equation}
    a_x=0,a_y=0,a_s=-\frac{1}{2}k(y^2-x^2)\, ,
\end{equation}
where $k=\frac{q_0}{P_0}g$ is the normalised quadrupole gradient which has the unit of $m^2$ and $g$ is the quadrupole gradient. The following Hamiltonian ca be derived to describe the quadrupole in thin-lens approximation
\begin{equation}
    H\approx H_D + \frac{1}{2}\tilde{\delta}(s-s_0)L\chi k(x^2-y^2)
\end{equation}
where $H_D$ is the approximated Hamiltonian for a drift space. The tracking maps for a thin-lens quadrupole are given by

\begin{equation}
\begin{pmatrix}
x^{F}\\y^{F}\\z^{F}
\end{pmatrix}
=
\begin{pmatrix}
x^{I}\\y^{I}\\z^{I}
\end{pmatrix}
\end{equation}
\begin{equation}
\begin{pmatrix}
p_x^{F}\\p_y^{F}\\p_z^F
\end{pmatrix}
=
\begin{pmatrix}
p_x^{I}-\chi k L x^I\\p_y^{I}+\chi k L y^I\\p_{z}^{I}
\end{pmatrix}
\end{equation}

This transfer map corresponds to a focusing lens in horizontal and defocusing lens in vertical direction. The transformation of $x'$ and $y'$ is given by
\begin{equation}
\begin{pmatrix}
(x')^{F}\\(y')^{F}
\end{pmatrix}
=
\begin{pmatrix}
(x')^{I}-k L x^I\frac{\chi}{1+\delta}\\(y')^{I}+k L y^I \frac{\chi}{1+\delta}
\end{pmatrix}
\end{equation}

\subsection{Accelerating RF Cavity}

The energy gain $\Delta E$ of a particle in an accelerating cavity with wave number $k = \frac{\omega}{c} = 2 \pi f$ can be \mbox{approximated by}
\begin{align}
  \Delta E = Z e  U \, \sin \left(\phi - k \, \frac{\sigma}{\beta_0}\right) \, ,
  \label{deltaE}
\end{align}
where $U$ is the average voltage during the particle's passage through the cavity~\cite{wolski}. In the approximation of a thin cavity, the following vector potential can be derived
\begin{align}
A_x = A_y =0 \, \quad A_s = - \frac{U}{\omega} \cos \left( \phi - k \frac{\sigma}{\beta_0}  \right) \, \tilde{\delta}(s) \, ,
\end{align}
where $\tilde{\delta}(s)$ is the Dirac function. Using the substitution $U_n = \frac{Z_0 e}{P_0 c} U$, the transfer map for $p_z$ can be deduced. 

The resulting expanded Hamiltonian for a thin cavity is then given by
\begin{equation}
    H\approx H_D+\chi q_0 \frac{UC}{\beta_0^2 E_0 2\pi h} \cos \left( \frac{2\pi h}{C}\sigma+\phi \right) L\tilde{\delta}(s-s_0)  \, .
\end{equation}

The thin-lens tracking maps for the accelerating RF cavity are the following
\begin{equation}
\begin{pmatrix}
x^{F}\\y^{F}\\z^{F}
\end{pmatrix}
=
\begin{pmatrix}
x^{I}\\y^{I}\\z^{I}
\end{pmatrix}
\end{equation}
\begin{equation}
\begin{pmatrix}
p_x^{F}\\p_y^{F}\\p_z^F
\end{pmatrix}
=
\begin{pmatrix}
p_x^{I}\\p_y^{I}\\p_{z}^{I}+\chi q_0\frac{1}{\beta_0^2}\frac{U}{E_0}L \sin(\frac{2\pi h}{C}+\phi)
\end{pmatrix}
\end{equation}
The change in $p_{\sigma}$ is, as expected, proportional to $q\frac{m_0}{m}$ by the relation between $p_{\sigma}$ and $E$, the energy transfer by the RF cavity of length $L$ corresponds to the expression given in Eq.~\eqref{deltaE} and yields
\begin{equation}
    \delta E=qUL\sin \left( \frac{2\pi h}{C} + \phi \right)
\end{equation}
\subsection{Thin Multipole}
Higher order magnetic fields are described in a more generic way as
\begin{equation}
B_y+iB_x=\sum_{n=1}^\infty (b_n+ia_n) \left( \frac{(x+iy)^n}{r_0^{n-1}}\right) \, .
\label{mfield}
\end{equation}
In this context, $n$ is the multiple order, $b_n$, $a_n$ are the multiple coefficients, one component is for the upright fields and the other one is for the slanted ones, which describe the field orientation for the contribution of each multiple order. The quantity $r_0$ is a reference radius. The magnetic field described in Eq.~\eqref{mfield} corresponds to the following vector potential
\begin{equation}
A_x=0, A_y=0, A_z=-Re\sum_{n=1}^\infty (b_n+ia_n) \left( \frac{x+iy}{r_0}\right)^{n-1} \, .
\end{equation}
Inserting this vector potential into the Hamiltonian in thin-lens approximation yields
\begin{equation}
    H\approx H_D-\frac{q_0}{P_0}\chi L \tilde{\delta}(s-s_0) Re\sum_{n=1}^\infty (b_n+ia_n) \left( \frac{x+iy}{r_0}\right)^{n-1} \, .
\end{equation}
The tracking maps for the thin multiple kick
\begin{equation}
\begin{pmatrix}
x^{F}\\y^{F}\\z^{F}
\end{pmatrix}
=
\begin{pmatrix}
x^{I}\\y^{I}\\z^{I}
\end{pmatrix}
\end{equation}
\begin{equation}
\begin{pmatrix}
p_x^{F}\\p_y^{F}\\p_z^F
\end{pmatrix}
=
\end{equation}
\begin{equation}
\begin{pmatrix}
p_x^{I}-\chi L Re \left [ \sum_{n=1}^\infty (k_n+i\hat{k_n})(x+iy)^{n-1}) \right ]\\p_y^{I}-\chi L Re \left [ \sum_{n=1}^\infty (k_n+i\hat{k_n})(x+iy)^{n-1}) \right ]\\p_{z}^{I}
\end{pmatrix}
\end{equation}

where $k_n$ and $\hat{k_n}$ are defined as:
\begin{equation}
    k_n=\frac{q_0}{P_0}\frac{a_n}{r_0^{r-}} ~~and~~ \hat{k_n}=\frac{q_0}{P_0}\frac{b_n}{r_0^{n-1}}
\end{equation}
\newpage

%
\end{document}